\documentclass[manuscript]{acmart}
\usepackage{placeins}
\usepackage{tikz}
\usetikzlibrary{arrows.meta,calc,positioning}
\DeclareRobustCommand{\MindBridge}{%
  {\fontencoding{T1}\fontfamily{qzc}\fontseries{m}\fontshape{it}\selectfont MindBridge}}
\makeatletter
\if@ACM@manuscript
  \newenvironment{mbwidetable}{\begin{table}[H]}{\end{table}}
\else
  \newenvironment{mbwidetable}{\begin{table*}[t]}{\end{table*}}
\fi
\makeatother
\AtBeginDocument{%
  }

\setcopyright{acmlicensed}
\copyrightyear{2018}
\acmYear{2018}
\acmDOI{XXXXXXX.XXXXXXX}
\acmConference[Conference acronym 'XX]{Make sure to enter the correct
  conference title from your rights confirmation email}{June 03--05,
  2018}{Woodstock, NY}
\acmISBN{978-1-4503-XXXX-X/2018/06}

\begin{document}

\title{\MindBridge{}: A Multilingual and Multimodal Tool for Interactive
Cognitive Monitoring}

\author{Xiaojie Gu}
\affiliation{%
  \institution{The University of Tokyo}
  \city{Tokyo}
  \country{Japan}}
\email{peettherapynoys@gmail.com}

\author{Haoyu Zhang}
\affiliation{%
  \institution{The University of Tokyo}
  \city{Tokyo}
  \country{Japan}}
\email{haoyu.zhang@weblab.t.u-tokyo.ac.jp}

\author{Qian Niu}
\affiliation{%
  \institution{The University of Tokyo}
  \city{Tokyo}
  \country{Japan}}
\email{qian.niu@weblab.t.u-tokyo.ac.jp}

\author{Fan Gao}
\affiliation{%
  \institution{The University of Tokyo}
  \city{Tokyo}
  \country{Japan}}
\email{fangao0802@gmail.com}

\author{Rio Yang}
\affiliation{%
  \institution{The University of Tokyo}
  \city{Tokyo}
  \country{Japan}}
\email{chocogeargo@gmail.com}

\author{Eleni Friligkou}
\affiliation{%
  \institution{Yale School of Medicine}
  \city{New Haven}
  \state{Connecticut}
  \country{United States}}
\email{eleni.friligkou@gmail.com}

\author{Kevin Jin}
\affiliation{%
  \institution{Yale School of Medicine}
  \city{New Haven}
  \state{Connecticut}
  \country{United States}}
\email{kevin.jin@yale.edu}

\author{Yutaka Matsuo}
\affiliation{%
  \institution{The University of Tokyo}
  \city{Bunkyo-ku}
  \state{Tokyo}
  \country{Japan}}
\email{matsuo@weblab.t.u-tokyo.ac.jp}

\author{Irene Li}
\affiliation{%
  \institution{The University of Tokyo}
  \city{Tokyo}
  \country{Japan}}
\email{irene.li@weblab.t.u-tokyo.ac.jp}

\renewcommand{\shortauthors}{Gu et al.}
\authorsaddresses{}


\begin{abstract}
We present \MindBridge{}, a clinician-informed, multilingual, multimodal cognitive
monitoring tool combining participant-controlled interviews with explainable
reports. Reports link lexical diversity and pause patterns to speech or text
evidence and disclose data quality and missingness. In a formative study of
typed interaction with 10 adults (seven aged 18--59; three aged 60 or older),
follow-ups prompted elaboration but sometimes introduced assumptions requiring
correction. Nine participants ranked interpretation-first reports highest; all
identified score-only reports as most likely to encourage fixation and
overreaction. On a fixed DementiaBank manifest, the best configuration achieved
70.3\% zero-shot accuracy on valid predictions versus a 56.0\% majority-class
baseline on the full manifest. Some configurations predicted only one category;
tests with generated samples revealed different error patterns across input
types and languages. 
\MindBridge{} makes evidence collection, analysis, and interpretation inspectable, linking participant control to the scope and limits of monitoring results.
\end{abstract}




\maketitle

\section{Introduction}

Cognitive health affects how people communicate, manage everyday activities,
and participate in family and community life. Connected discourse, such as
describing a picture or recounting a day, offers a window into these abilities
because it draws on lexical retrieval, semantic knowledge, memory,
narrative organization, executive control, and speech production
\cite{taler2008language,boschi2017connected,mueller2018connected}. Computational
approaches offer a promising means of examining how these abilities manifest in
discourse, from analyzing linguistic patterns in transcripts and temporal and
acoustic patterns in speech
\cite{petti2020systematic,delafuentegarcia2020artificial} to extracting transcript
features with large language models (LLMs) \cite{heitz2025linguistic} and
evaluating text-only and audio-text LLMs \cite{taherinezhad2026adaptation}.
These capabilities create opportunities for more accessible and multilingual cognitive research,
but model performance alone does not explain how the evidence is produced or
what a result can reasonably mean.

Existing conversational systems for cognitive assessment use backchannels and
scaffolding to support engagement and task understanding
\cite{ding2022talktive,hu2024scaffolding}. Human-centered explainability
research examines how result interfaces support understanding and appropriate
reliance \cite{liao2020questioning,bucinca2021trust}. In cognitive monitoring,
interaction therefore matters both before and after model analysis. Beyond
task completion, participants need to be able to express their intended
meaning, correct misunderstandings, and decide what to share. A follow-up may
help them elaborate, but it may also require them to correct an assumption
before continuing, changing the response later used for analysis.
After analysis, the presentation of scores, uncertainty, and explanations
shapes how readers understand the output
\cite{lipkus2007numeric,kay2016bus,hullman2019pursuit}.

The challenge is to explain not only a model's output but also how its
supporting evidence was collected and analyzed. Input scope and
modality-specific processing affect which information reaches the summary.
Connecting these stages extends human-centered explainability to how
monitoring evidence is formed and what conclusions it can support.

To investigate this design space, we present \MindBridge{}, a clinician-informed,
multilingual, and multimodal cognitive monitoring tool.
Structured input from a psychiatry resident informed the system's evidence
priorities, confound disclosures, and result presentation.
Through a bounded adaptive interview, the system supports elaboration and
offers participants explicit controls over what enters the record. Rather than
treating the resulting sample as a finished input, it preserves the prompts,
responses, and control actions that shaped its speech and text evidence. The
same evidence trail continues into a researcher workbench, where observations
remain tied to their source modality, data quality, and missingness. Together,
the workflow makes visible how interaction forms evidence, how multimodal
analysis transforms it, and how that evidence supports an interpretation.

Our evaluation addresses three questions:

\begin{description}
  \item[RQ1:] How do bounded adaptive follow-ups and task-level controls
  (rereading and prompt redirection) change what participants contribute to
  the record, and what control do they retain over it?
  \item[RQ2:] How do people interpret and rely on the same sample-specific
  evidence when it is presented as a bare score, a five-dimension breakdown,
  or an interpretation-first report?
  \item[RQ3:] How does \MindBridge{}'s multimodal analysis pipeline behave under
  fixed recording and controlled stress tests, and what limitations associated
  with class, modality, and language can aggregate performance obscure?
\end{description}

\begin{figure*}[t]
  \centering
  \includegraphics[width=\textwidth]{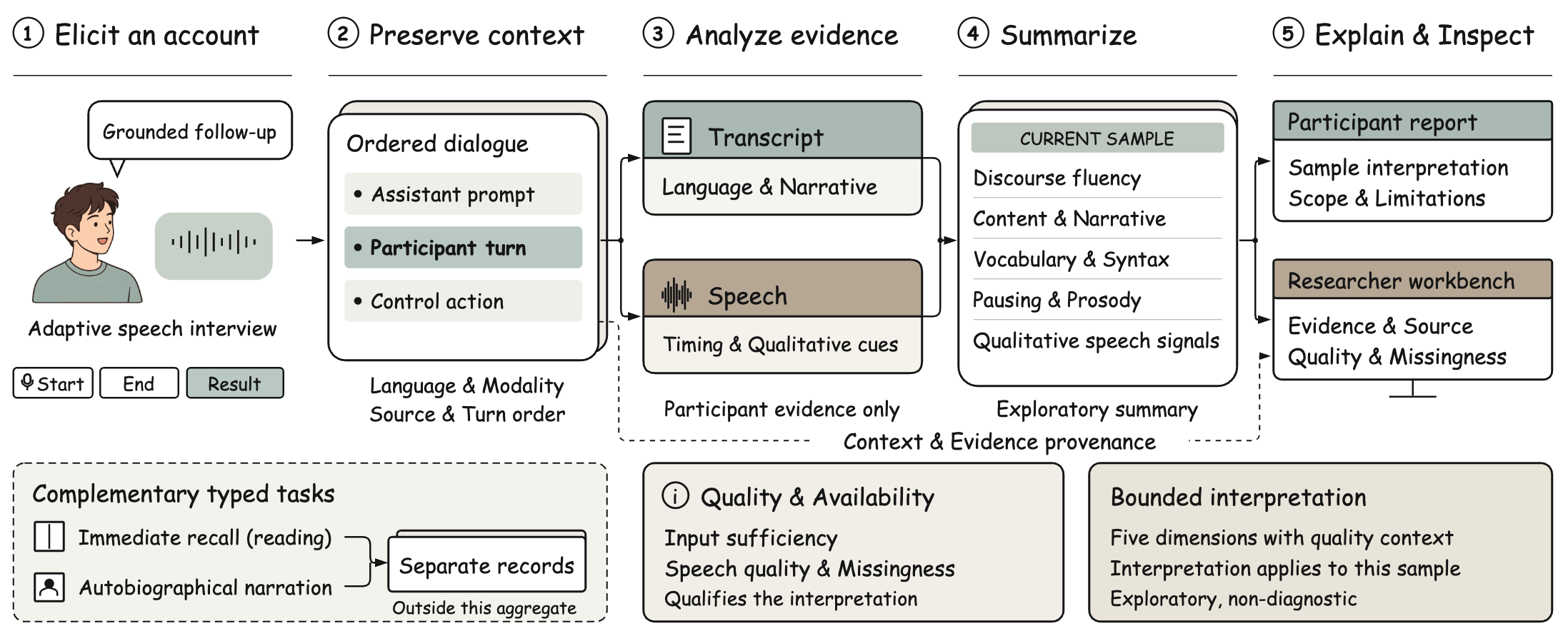}
  \caption{The \MindBridge{} cognitive monitoring workflow. Steps~1 and~2
  elicit a spoken account and preserve the prompts, responses, and participant
  controls. Steps~3 and~4 analyze transcript and speech evidence and summarize
  the current sample across five dimensions. Step~5 presents the participant
  report and researcher workbench. The two complementary typed tasks produce
  separate records for RQ1.}
  \Description{A left-to-right workflow presents five numbered steps: elicit an
  account, preserve context, analyze evidence, summarize, and explain and
  inspect. A participant-controlled adaptive speech interview produces an
  ordered record of assistant prompts, participant turns, and control actions.
  Participant evidence branches into transcript and speech analysis before
  forming a five-dimension summary. The summary leads to a participant report
  explaining the sample and its limits and a researcher workbench showing
  evidence sources, quality, and missingness. A dashed path carries context and
  evidence provenance to the workbench. A separate panel shows the two
  complementary typed tasks producing records outside this aggregate. Lower
  panels emphasize data quality, availability, and sample-bounded,
  exploratory interpretation.}
  \label{fig:monitoring-workflow}
\end{figure*}

Our evaluation examines the workflow in
Figure~\ref{fig:monitoring-workflow} at three points. Ten adults completed
a typed adaptive interview and two complementary typed tasks for RQ1, interpreted
two fictional cases, and
compared three report formats for RQ2.
For RQ3, we applied \MindBridge{}'s analysis workflow under changes in input scope,
modality, and language using 109 English DementiaBank recordings
\cite{becker1994natural}, 20 controlled older-adult personas at four programmed
communication-difficulty levels, and Chinese and Japanese cross-lingual stress
tests. This design connects the two human-facing boundaries while using the
technical checks to locate failures within the analysis pathway.

The user study highlights the need to support elaboration while preserving
participants' ability to correct the interviewer and control what they share.
Participants favored reports that explain the evidence and its limits before
presenting a score. They saw the score-only report as the most likely to draw
attention away from the evidence behind the number.
The technical checks expose complementary failures. Using the
same workflow, the best fixed-recording configuration reached 70.3\%
zero-shot accuracy, while several configurations collapsed toward one class.
Exact synthetic-level recovery fell from 85.0\% with text to 26.25\% with speech and
50.0\% with speech plus text, while identical Chinese and Japanese aggregate
accuracy masked sharply different sensitivity and specificity. Together, these
findings show that accountable interpretation requires visible evidence
transformations, not merely an explanation of the final score.

This work makes four contributions:

%
%
%

\begin{itemize}
  \item We present the design and implementation of \MindBridge{}, a
  clinician-informed, multilingual, and multimodal cognitive monitoring tool.
  It connects bounded adaptive speech interviewing and participant control
  with participant and researcher views of transcript and speech evidence,
  data quality, and missingness. Complementary immediate recall and
  autobiographical narration tasks are recorded separately.

  \item We articulate a process-oriented design account of human-centered
  explainability in cognitive monitoring, grounded in \MindBridge{} and its
  evaluations. Explanations should connect how a sample was elicited, which
  evidence entered the analysis, and what conclusions the result can support,
  rather than justify only the final score.

  \item We report empirical findings from a formative study of typed
  interaction with 10 adults. Participants reported adding unplanned details
  and identified follow-ups that assumed events or feelings they had not described.
  They favored interpretation-first reports, yet some still inferred overall cognitive
  ability from sample-specific evidence. These findings inform design
  implications for participant control and evidence-first reporting.

  \item We empirically characterize \MindBridge{}'s analysis workflow,
  identifying configurations that assigned all recordings to one class,
  differences between text and speech in recovering synthetic difficulty
  levels, and error patterns hidden
  by equal accuracy across translated and synthesized language versions.
  On a fixed DementiaBank manifest, the best-performing configuration achieved
  70.3\% zero-shot accuracy on valid predictions, compared with a 56.0\%
  majority-class baseline on the full manifest.
\end{itemize}

\section{Related Work}

\subsection{Conversational Systems for Cognitive Assessment}

Speech and language provide established evidence for cognitive assessment
through lexical choice, discourse organization, fluency, prosody, and
articulation \cite{taler2008language,boschi2017connected,mueller2018connected,
fraser2016linguistic}. These signals depend on the elicitation setting because
picture description, narration, interviews, and spontaneous conversation place
different demands on participants \cite{boschi2017connected,mueller2018connected}.
The resulting sample therefore reflects both participant expression and the
interaction through which it was produced.

Conversational agents make this interaction adaptive. Systems use backchannels
and scaffolding to support engagement and task understanding in conversational
cognitive assessment \cite{ding2022talktive,hu2024scaffolding}. Conversational
surveys have elicited more specific responses than conventional web forms
\cite{xiao2020tell}. Interviewer speech can also influence automatic assessment
results \cite{pereztoro2021influence}.

Speech and transcripts provide complementary evidence for multimodal analysis
\cite{petti2020systematic,delafuentegarcia2020artificial}. Recent LLM approaches
include transcript feature extraction \cite{heitz2025linguistic} and audio-text
modeling \cite{taherinezhad2026adaptation}. Systems such
as Winterlight, Canary Cognitive, and ki:elements' SB-C illustrate practical
speech-based screening and monitoring \cite{robin2021digital,canaryCognitive,
troger2022validation}; FRONTEO's KIBIT supports expert text analysis
\cite{fronteoKIBIT}, while related work examines neurocognitive disorder
classification from free-conversation transcripts \cite{horigome2022identifying}.
These systems demonstrate practical applications of speech and text analysis.
Our focus is on bounding interviewer influence, retaining participant control,
and preserving elicitation context as the sample enters speech and text analysis.

\subsection{Human-Centered Explainability}

Human-centered explainability asks whether people can understand an
AI-supported result, recognize its limits, and decide when to rely on it.
Relevant guidance emphasizes visible capabilities and uncertainty, as well as
explanations shaped by users' questions and goals
\cite{amershi2019guidelines,liao2020questioning}. In health-related settings,
numeric and visual formats influence risk interpretation
\cite{lipkus2007numeric}, while uncertainty displays remain dependent on task
and context \cite{kay2016bus,hullman2019pursuit}. Explanations alone do not
necessarily prevent overreliance \cite{bucinca2021trust}.

Interactive cognitive monitoring must also account for how prompts, repairs,
review choices, and input modality form the evidence, and how result order
shapes its interpretation. This creates a need to connect each claim to the
interaction, quality, missingness, and modality that support it. \MindBridge{}
examines this connection across elicitation, analysis, and result communication
rather than adding explanation only after a result has been produced.

\section{\MindBridge{}: An Interactive Multimodal Cognitive Monitoring Tool}
\label{sec:mindbridge}

\MindBridge{} is a web-based, clinician-informed, multilingual cognitive
monitoring tool that collects speech and text samples and summarizes observable
communication patterns. It is designed for cognitive monitoring research and
self-reflection rather than diagnosis or clinical assessment. The current
prototype produces an exploratory summary of the sample supplied in a session;
it does not determine a person's cognitive status and is not clinically
validated.

\begin{figure*}[htbp]
  \centering
  \begingroup
  \definecolor{mbInterview}{HTML}{6658CA}
  \definecolor{mbRecall}{HTML}{548AB2}
  \definecolor{mbNarration}{HTML}{8C72AE}
  \definecolor{mbReport}{HTML}{388681}
  \tikzset{mb frame/.style={}}
  \newcommand{\mbShot}[2]{%
    \begin{tikzpicture}
      \node[mb frame, inner sep=3pt, outer sep=0pt, rounded corners=3pt,
        fill=white, draw=#1!48!white, line width=.45pt] (shot)
        {#2};
      \draw[#1,line width=1.1pt,line cap=round]
        ([xshift=4pt,yshift=-1.5pt]shot.north west) --
        ([xshift=-4pt,yshift=-1.5pt]shot.north east);
    \end{tikzpicture}%
  }
  \newcommand{\mbPanelTitle}[1]{\par\vspace{2pt}{\small\bfseries #1\par}}
  \newcommand{\mbRowTitle}[1]{{\raggedright\sffamily\bfseries\small #1\par}\vspace{3pt}}
  \mbRowTitle{Interview and typed tasks}
  \begingroup
  \tikzset{mb frame/.style={minimum height=1.045\linewidth}}
  \begin{minipage}[t]{.324\linewidth}
    \vspace{0pt}\centering
    \mbShot{mbInterview}{%
      \includegraphics[width=\dimexpr\linewidth-7pt\relax,trim=250bp 0bp 255bp 100bp,clip]{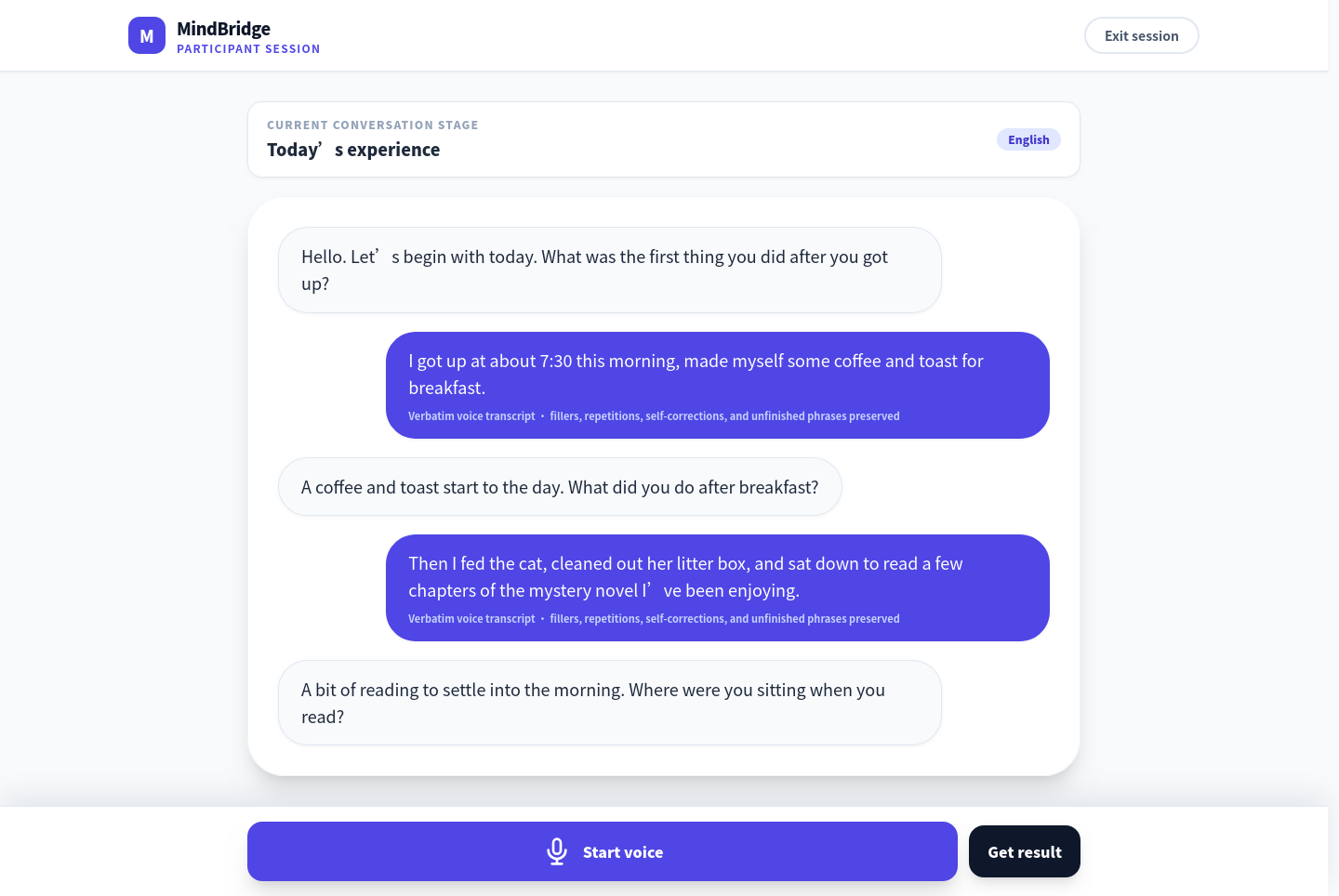}}
    \mbPanelTitle{(a) Adaptive interview}
  \end{minipage}\hfill
  \begin{minipage}[t]{.324\linewidth}
    \vspace{0pt}\centering
    \mbShot{mbRecall}{%
      \includegraphics[width=\dimexpr\linewidth-7pt\relax]{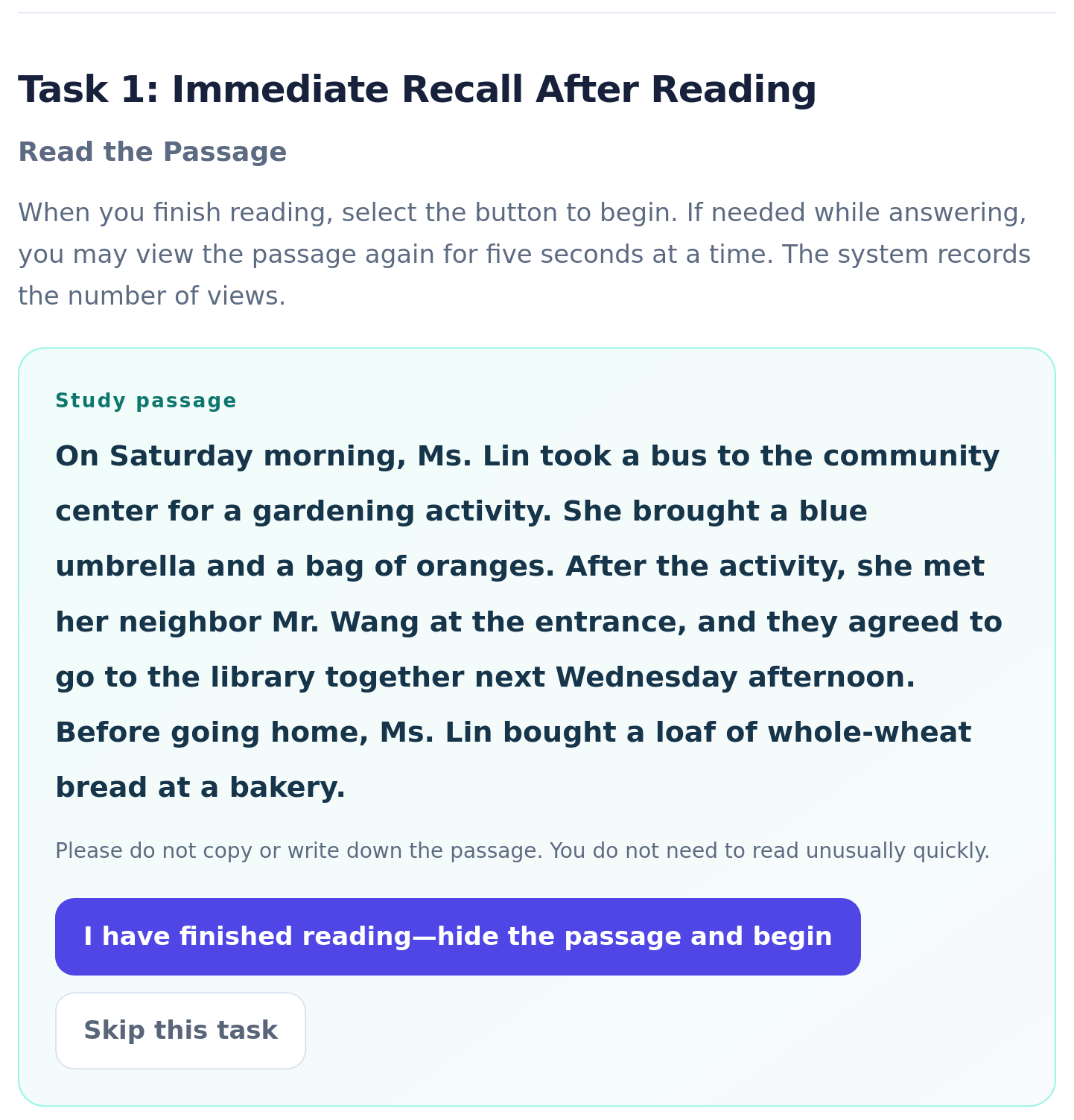}}
    \mbPanelTitle{(b) Immediate recall}
  \end{minipage}\hfill
  \begin{minipage}[t]{.324\linewidth}
    \vspace{0pt}\centering
    \mbShot{mbNarration}{%
      \includegraphics[width=\dimexpr\linewidth-7pt\relax]{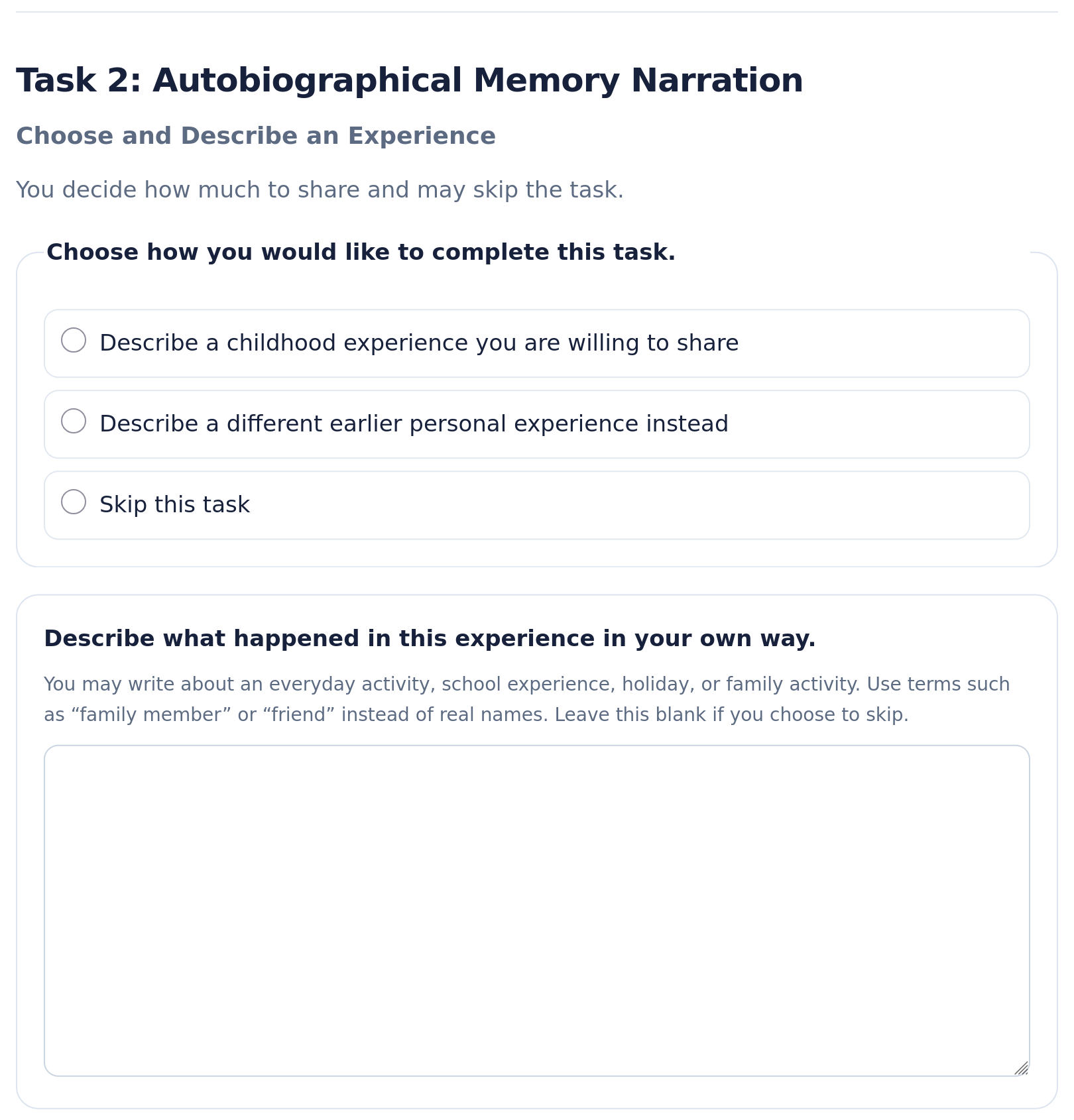}}
    \mbPanelTitle{(c) Autobiographical narration}
  \end{minipage}\par
  \endgroup
  \vspace{8pt}
  \mbRowTitle{Report packages for Case H}
  \begin{minipage}[t]{.324\linewidth}
    \vspace{0pt}\centering
    \mbShot{mbReport}{%
      \includegraphics[width=\dimexpr\linewidth-7pt\relax]{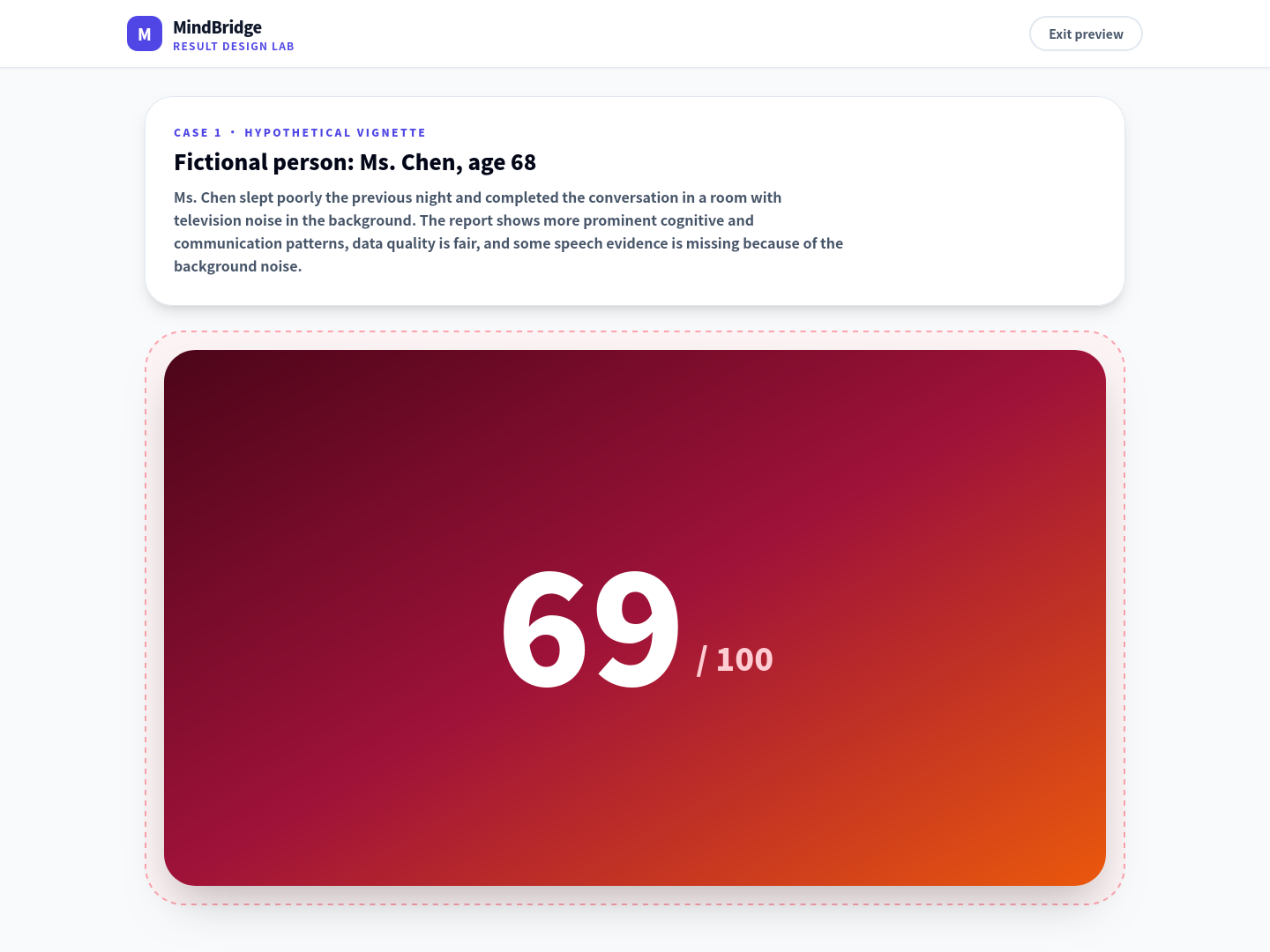}}
    \mbPanelTitle{(d) Score-only report}
    \vspace{12pt}
    {\fontsize{20}{22}\selectfont\color{mbInterview}\MindBridge{}\par}
    {\sffamily\fontsize{7.5}{9}\selectfont Participant-controlled cognitive monitor\par}
  \end{minipage}\hfill
  \begin{minipage}[t]{.324\linewidth}
    \vspace{0pt}\centering
    \mbShot{mbReport}{%
      \includegraphics[width=\dimexpr\linewidth-7pt\relax]{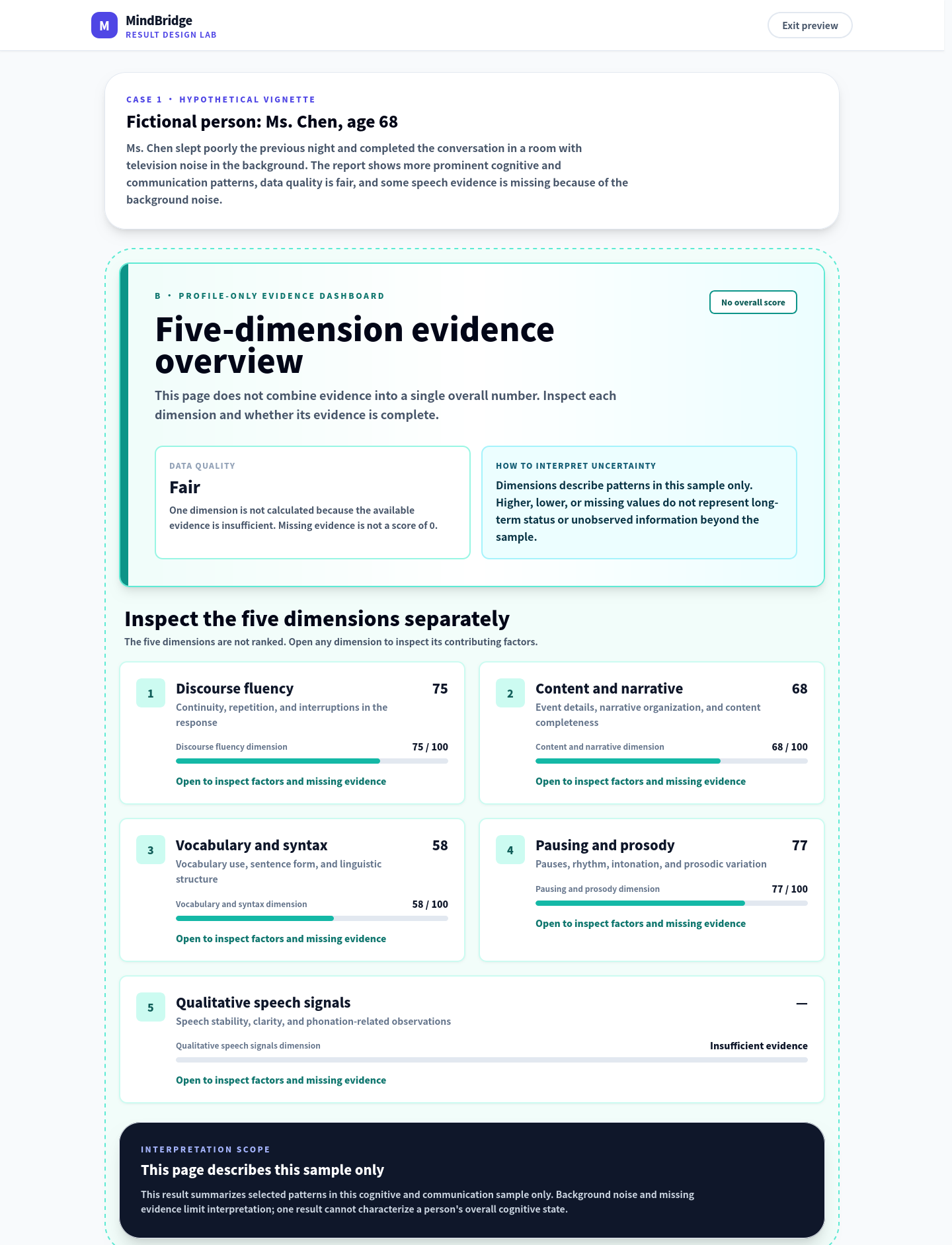}}
    \mbPanelTitle{(e) Five-dimension report}
  \end{minipage}\hfill
  \begin{minipage}[t]{.324\linewidth}
    \vspace{0pt}\centering
    \mbShot{mbReport}{%
      \includegraphics[width=\dimexpr\linewidth-7pt\relax]{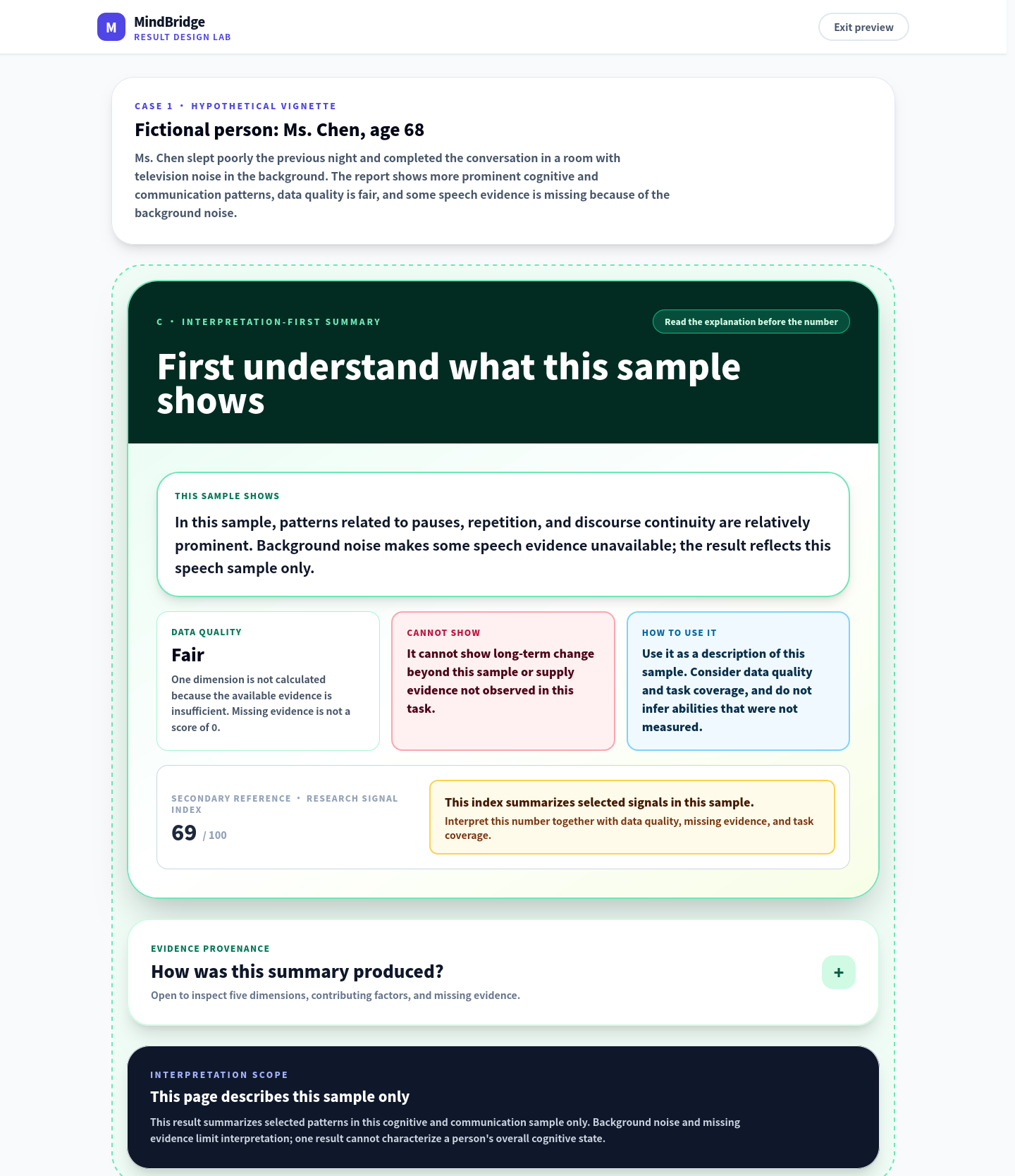}}
    \mbPanelTitle{(f) Interpretation-first report}
  \end{minipage}\par
  \endgroup
  \caption{\MindBridge{}'s elicitation interfaces and the report packages compared
  in the user study. (a): In the speech prototype, participants control recording
  and respond to follow-ups. (b): They read a passage before reconstructing it.
  (c): They choose a personal experience to narrate or skip the task.
  Panels (d--f) display the same Case~H vignette above each result panel.
  (d): Package~A's result panel shows only the overall index.
  (e): Package~B presents five dimensions and expandable evidence.
  (f): Package~C explains what the sample
  shows and its limits before presenting the index.}
  \Description{Six English screenshots in two rows. Across the top, the
  interview shows two speech transcripts, follow-up questions, and recording
  and result buttons. The recall task shows a passage with controls to hide
  it and begin answering or skip. The narration task offers a childhood
  experience, another earlier experience, or a skip option above a text field.
  Across the bottom, each report opens with the same description of background
  noise, fair quality, and missing speech evidence. The first result panel centers
  a score of 69 out of 100. The second shows fair data quality and five
  dimension cards, one marked insufficient evidence. The third presents the
  interpretation and limits above a secondary score of 69 and a collapsed
  evidence control.}
  \label{fig:participant-interfaces}
\end{figure*}

The workflow in Figure~\ref{fig:monitoring-workflow} includes a bounded adaptive speech interview plus two
linked typed modules: immediate recall after reading and autobiographical
narration (Figure~\ref{fig:participant-interfaces}(a--c)). The interview supplies
the sample analyzed for the participant
report. The typed modules preserve task-specific responses and participant
control actions for RQ1, but are stored separately and do not enter the current
multimodal aggregate. When a participant requests a result, \MindBridge{} checks
whether the interview input is sufficient, analyzes transcript and speech
evidence, and organizes the available observations into five participant-facing
dimensions derived from the reviewed feature registry. Data quality and
missingness are reported separately to qualify the interpretation. The
participant view explains what the current sample shows and what it cannot
establish. A linked researcher workbench exposes the ordered dialogue, input
modality, data quality, missingness, feature evidence, and analysis versions.

\MindBridge{} supports human-centered explainability across this monitoring
workflow by making elicitation, analysis, and result communication traceable.
During elicitation, \textbf{bounded adaptivity} instructs the model to ground
follow-ups in participant-provided details and preserves each prompt so that
departures from this constraint remain inspectable. \textbf{Participant-controlled
turn taking} lets the participant decide when a contribution begins and ends,
while \textbf{legible recovery} exposes system states and retry options. During
analysis and result communication, \textbf{evidence provenance} keeps
participant speech, assistant prompts, transcripts, speech measurements, and
model observations distinguishable, while \textbf{scope-first communication}
carries source, quality, and missingness into the result view. Together, these
mechanisms make visible how each sample was elicited, transformed, and
summarized.

The remainder of this section describes the monitoring scope and evidence
model (Section~\ref{sec:monitoring-scope}), the bounded adaptive interview
(Section~\ref{sec:bounded-interview}), complementary memory and narrative tasks
(Section~\ref{sec:complementary-tasks}), and multimodal analysis and result
communication (Section~\ref{sec:multimodal-analysis}).
The clinical collaboration described below informed the scope of
the monitoring claims, the evidence vocabulary, feature priorities, confound
warnings, and participant presentation. The technical checks in
Section~\ref{sec:technical-boundary-checks} test how this implemented workflow
behaves across input modalities, classes, and languages.

\subsection{Defining the Monitoring Scope and Evidence Model}
\label{sec:monitoring-scope}

To determine what a speech and language monitoring tool can responsibly report,
we sought clinical input before implementing the evidence model. An initial design
consultation involved a psychiatry resident physician (MD, MSc) with experience
in clinical practice, neuroscience, and health data research. The physician
emphasized that speech is only one component of dementia assessment.
Clinical interpretation also depends on longitudinal cognitive decline, change
in daily functioning, patient and family history, cognitive screening,
behavioral and affective observation, and alternative explanations such as age,
education, language, hearing, mood, medication, fatigue, and comorbidity.
\MindBridge{} therefore treats speech and language patterns as sample-specific
monitoring evidence rather than a standalone cognitive assessment. This scope
also requires the interface to keep potential confounds and evidence limits
visible.

We then compiled an inventory of 11 speech and language feature families for
the monitoring evidence model and sent it to the physician for asynchronous
structured annotation.
The packet described binary corpus classification, disclosed the absence of patient
history, and included labeled speech examples. After listening, the physician
made three structured judgments for each feature family, yielding 33 judgments
that cover clinical-exam relevance, direct MoCA/MMSE coverage
\cite{nasreddine2005moca,folstein1975mmse}, and expected computational yield.
The physician supplied written rationales for every Maybe or No judgment and,
after considering the inventory and recordings, identified no additional
feature family that was clearly missing (Table~\ref{tab:clinical-feature-review}).
After implementation, the physician reviewed the prototype UI containing
the participant-facing dimensions and summary. Thus, clinical involvement
spanned problem framing, feature prioritization, sample speech review, and
interface review. This collaboration guided the monitoring design but does not
constitute clinical validation.

\begin{table*}[t]
  \caption{Structured annotation of 11 speech and language feature families by one
  psychiatry resident physician. Exam and Screen reproduce the physician's judgments about
  clinical-exam relevance and direct MoCA/MMSE coverage. Yield denotes
  anticipated analytic value. The final column records
  our subsequent design translation.}
  \label{tab:clinical-feature-review}
  \centering
  \scriptsize
  \setlength{\tabcolsep}{3.5pt}
  \renewcommand{\arraystretch}{1.08}
  \begin{tabular}{@{}p{0.205\linewidth}ccc p{0.43\linewidth}@{}}
    \toprule
    Feature family & Exam & Screen & Yield & Review rationale and design translation \\
    \midrule
    Lexical diversity & No & No & Maybe & Difficult to assess manually but potentially informative computationally; supporting registry feature. \\
    Syntactic complexity & No & No & Maybe & Difficult to assess manually but potentially informative computationally; supporting registry feature. \\
    Discourse fluency & Yes & Yes & Yes & Routinely clinically salient; core registry feature. \\
    Pause and speech rhythm & Yes & No & Maybe & Clinically informative but potentially confounded by depression; supporting feature with a confound disclosure. \\
    Prosody & Yes & No & Maybe & Clinically informative but potentially confounded by depression or psychosis; supporting feature with a confound disclosure. \\
    Generic acoustic characteristics & No & No & No & Potentially confounded by age, race, and socioeconomic status; omitted from the participant-facing aggregate. \\
    Speech disfluency & Yes & Yes & Yes & Routinely clinically salient; core registry feature. \\
    Linguistic structure & No & No & Maybe & Speech rate is clinically observable, whereas other structures are difficult to quantify manually; depression may confound rate; supporting feature. \\
    Content quality & Yes & No & Yes & Clinically salient completeness and relevance of the account; core registry feature. \\
    Higher-level text features & Yes & No & Yes & Clinically salient global characteristics of the account; core registry feature. \\
    Clinical acoustic biomarkers & No & No & Maybe & Difficult to assess manually but potentially informative computationally; retained as supporting qualitative speech evidence and presented as Qualitative speech signals; no clinical biomarker claim is made. \\
    \bottomrule
  \end{tabular}
\end{table*}

The physician supplied ordinal judgments establish the priorities in the
feature registry. A Yes judgment makes a family eligible for core evidence,
Maybe restricts it to supporting evidence, and No excludes it from the
participant-facing aggregate. The registry also retains the corresponding
rationale and confound warning. To operationalize this ordering, we use fixed
heuristic weights of 1 for the four Yes families and 0.5 for the six Maybe
families. The categorical priorities and rationales come from the physician;
their numeric conversion is our design choice. We organize the ten
retained families into five participant-facing dimensions: discourse fluency,
content and narrative, vocabulary and syntax, pausing and prosody, and
qualitative speech signals. The physician subsequently reviewed their
presentation in the prototype interface.

The annotations produced three requirements for the monitoring design. First,
measures that are difficult for a person to assess but easy to compute must
remain source-labeled rather than inherit clinical authority from
automation. Second, features susceptible to psychiatric or demographic
confounding require visible limitations and should not dominate an aggregate.
Third, absence from MoCA or MMSE does not imply that a feature lacks research
value; instead, the interface should distinguish routine clinical observation,
computational evidence, and uncertainty. In the implementation, unavailable
evidence remains null and is removed from the denominator rather than treated
as an observed zero. Model-derived observations supply feature-level evidence,
while a deterministic rule computes dimension and overall summaries from the
versioned registry. This separation makes it possible to distinguish the
contribution of clinical prioritization, model observations, and deterministic
aggregation to a result.

\subsection{Collecting Participant-Specific Accounts Through a Bounded Adaptive Interview}
\label{sec:bounded-interview}

The adaptive interview collects speech about the participant's activities
on the day of the session. While all participants receive the same opening
question, follow-ups adapt to the events and details each participant
introduces. At entry, a participant enters their own ID, selects Chinese,
English, or Japanese, and reviews an online consent page. The system creates
a session only after the participant selects agreement. The opening question,
presented in the selected language, asks about the participant's first activity
after getting up that day. This topic can
elicit temporal, spatial, social, and event details within a familiar narrative,
without introducing a sequence of unrelated cognitive tasks.

The interview proceeds through a repeated, participant-controlled loop
(Table~\ref{tab:mindbridge-interaction-loop}). The participant selects the
microphone control to begin a turn and selects it again to commit the turn.
During this process, the interface presents persistent status feedback for
connecting, ready-to-speak, receiving speech, waiting for a reply, and failure
states. Completed participant transcripts and assistant replies appear as
role-separated chat bubbles, allowing the participant to see the conversational
history and the current point in the narrative. The current prototype supports
participant speech input rather than fully speech-to-speech interaction:
participants speak, whereas the assistant's response is displayed as text.
Figure~\ref{fig:participant-interfaces}(a) shows the participant-facing state
after two illustrative speech turns. The fixed opening, labeled speech
transcripts, and one-question follow-ups remain visible as interaction context,
while the persistent controls leave initiation of the next turn or result
request with the participant.


\begin{table*}[t]
  \caption{The participant-controlled interaction loop and the records retained
  by \MindBridge{}.}
  \label{tab:mindbridge-interaction-loop}
  \centering
  \small
  \begin{tabular}{@{}p{0.14\linewidth}p{0.22\linewidth}p{0.33\linewidth}p{0.23\linewidth}@{}}
    \toprule
    Phase & Participant action & System response & Retained record \\
    \midrule
    Entry & Enters their ID, selects a language, reviews the consent page, and selects agreement & Creates a session and displays a fixed opening question & Participant-provided ID, language, and consent record \\
    Speaking & Starts the microphone, speaks, and ends the turn manually & Displays the current capture state and commits one bounded speech turn & Ordered speech segment and explicit turn boundary \\
    Follow-up & Reads the reply and decides what to describe next & Applies a prompt policy requiring one contextual question grounded in the participant's account and asynchronously displays the transcript & Participant transcript, assistant reply, role, sequence, and time \\
    Repair & Repeats a turn or retries after a visible failure & Uses a neutral repetition request and preserves the contribution during a visible retry & Repair request, repeated contribution, and turn order \\
    Completion & Requests a result or continues speaking & Checks data sufficiency and distinguishes saving from analysis & Data-quality status, evidence provenance, and report version \\
    \bottomrule
  \end{tabular}
\end{table*}

\MindBridge{} uses participant-controlled turn boundaries: speech is captured only
after the participant explicitly starts a turn and is committed only when the
participant ends it. The interviewer responds to the original speech and prior
dialogue, while a separate transcription branch retains a textual record. The
later monitoring analysis can use speech and transcript evidence without
conflating their sources. This architecture preserves an immediate
conversational loop and a traceable evidence record.
The deployed implementation uses gpt-realtime-2 for speech interaction and
qualitative speech observations. GPT-5.4 generates typed follow-ups, analyzes
stored transcripts, and performs the final structured evidence-fusion step. It
assigns feature-level scores, after which the server deterministically computes
the five dimension scores and overall research-signal index.
The technical evaluation reused this analysis workflow while varying model
configuration, input scope, modality, and language.

Because follow-up questions can change the sample that is later analyzed, the
adaptive policy is intentionally narrow. The interviewer remains on what
the participant does from waking until the present and can follow concrete
anchors that the participant introduces, such as an action, place, person,
object, journey, event order, or explicitly stated feeling. Each normal reply
uses one primary conversational move, is normally one or two short sentences,
and asks one low-burden question. The policy prohibits inferring an unstated
routine or emotion, praising the participant's performance, stacking several
questions, or introducing picture description, listening tests, or other
assessment tasks. If
a participant cannot remember or declines to answer, the interviewer moves to
a later point without pressure. If speech is unintelligible, it uses a neutral
request to repeat rather than guessing the content or an acoustic cause. These
prompt constraints aim to make adaptation responsive to the participant while
limiting how much the agent can reshape the elicitation task.

The participant may request completion at any time. Before producing a
monitoring result, the system checks that the session contains sufficient
spoken material and usable non-silent speech. This check prevents a sparse or
unusable recording from being presented as an interpretable result. If a
requirement is unmet, the interface explains what material is missing and lets
the participant continue. Once the conditions are met, the application supplies
a deterministic wrap-up while still allowing the participant to add more.
During completion, the interface distinguishes saving from analysis and offers
retry or safe exit choices after a failure.

\subsection{Collecting Complementary Memory and Narrative Samples}
\label{sec:complementary-tasks}

\begin{figure*}[t]
  \centering
  \includegraphics[width=\textwidth]{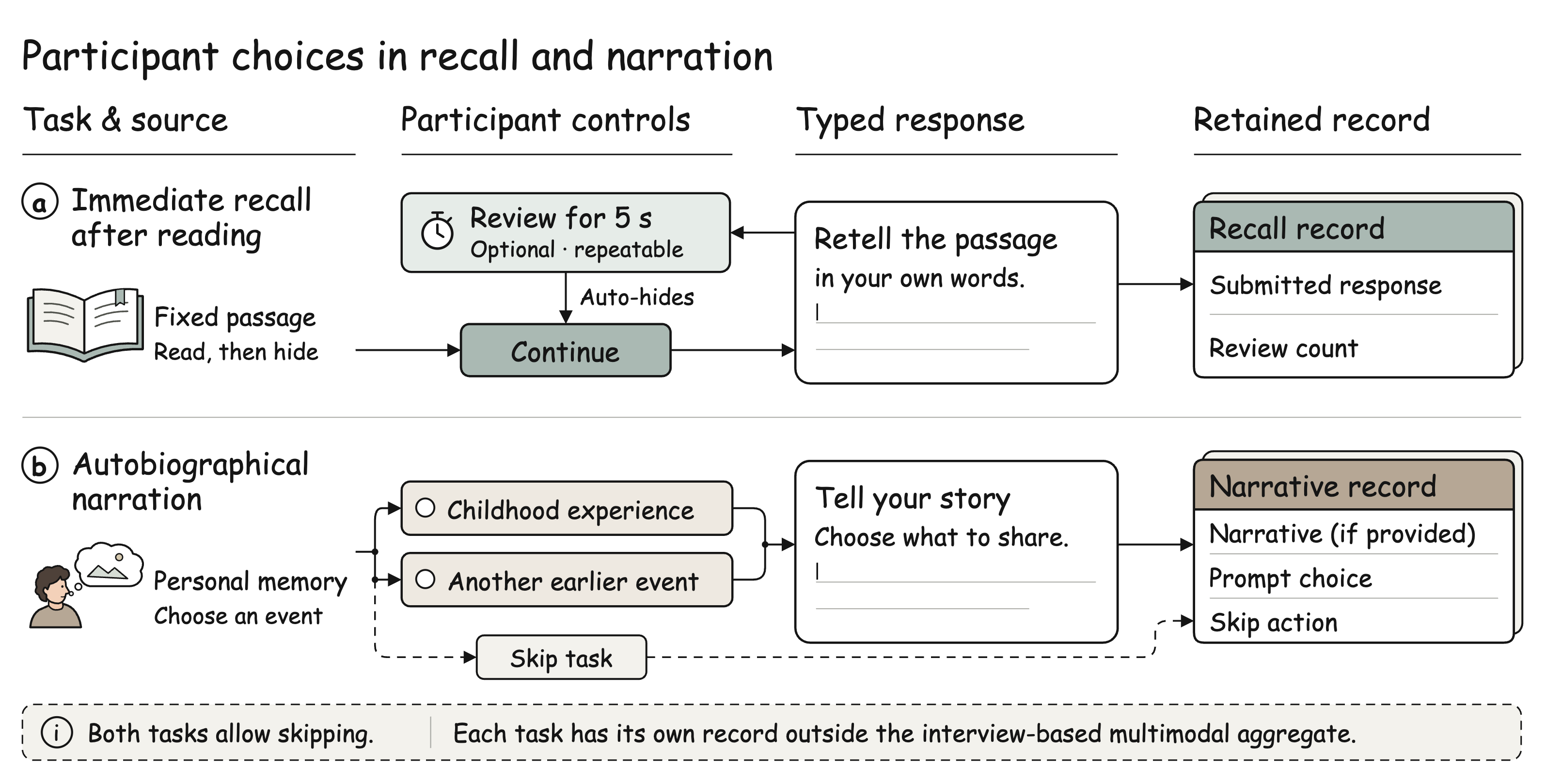}
  \caption{Two complementary typed tasks in \MindBridge{}, showing each task's
  source, participant controls, response, and retained record. (a): Immediate
  recall after reading elicits reconstruction of a fixed passage, with optional
  five-second reviews; the submitted response and review count are retained.
  (b): Autobiographical narration allows participants to choose a childhood or
  other earlier experience and decide what to share; the prompt choice, skip
  action, and narrative when provided are retained. Both tasks allow skipping
  and produce separate records outside the interview-based multimodal aggregate.}
  \Description{Two rows share four columns for task and source, participant
  controls, typed response, and retained record. In row (a), participants
  read and hide a fixed passage, then retell it in their own words. They may
  reopen the passage for five seconds at a time. The recall record contains
  the submitted response and review count. In row (b), participants choose
  a childhood experience or another earlier event and decide what to share.
  A dashed skip path bypasses the writing step. The narrative record contains
  the prompt choice, skip action, and narrative if provided. A note states
  that both tasks allow skipping and keep their own records outside the
  interview-based aggregate.}
  \label{fig:complementary-tasks}
\end{figure*}

A single conversational format provides only one view of participant
expression. We selected two complementary tasks: immediate recall after
reading and autobiographical memory narration. The physician subsequently
reviewed the clinical relevance of the memory and language features these tasks
were intended to elicit and considered them appropriate complements to the
adaptive interview. The tasks vary the source of memory and the available forms
of participant control (Figure~\ref{fig:complementary-tasks}). They are
standalone typed activities whose task-specific records support RQ1. They are
stored separately from the speech-input interview and do not contribute to the
current participant-facing aggregate.

Immediate recall after reading presents a short passage and asks participants
to reconstruct its content; the choice to use an optional five-second review is
recorded alongside the response rather than treated as an error.
Autobiographical memory narration invites a childhood or other earlier
experience, while participants choose which details to disclose and may skip
the task. These mechanisms retain the optional review choice as provenance and
make disclosure and skip control explicit at the point of response, rather than
presenting each task as a static response box. Together with the adaptive
interview, these tasks give the study workflow three complementary
elicitation pathways spanning recently presented information, personally
grounded recollection, and an unfolding everyday account.
In Figure~\ref{fig:participant-interfaces}, panel~(b) shows the passage and
reading controls for Task~1, and panel~(c) shows the prompt choices for Task~2.
Appendix~\ref{app:additional-interface-screens} provides larger
task views in Figures~\ref{fig:immediate-recall-screen}
and~\ref{fig:autobiographical-memory-screen}.

\subsection{Analyzing Multimodal Evidence and Communicating Monitoring Results}
\label{sec:multimodal-analysis}

After a participant requests a result and the quality requirements are met,
\MindBridge{} constructs a session-level evidence record from three sources:
deterministic temporal speech measures, qualitative observations of the speech
signal, and linguistic and narrative evidence derived from participant
transcripts.
Assistant turns are retained as elicitation context but are not treated as
participant evidence. Available observations are mapped to the
physician-annotated feature registry described above and organized into five
participant-facing dimensions: discourse fluency, content and narrative,
vocabulary and syntax, pausing and prosody, and qualitative speech signals.
Feature eligibility, priority, rationales, and confound disclosures remain
attached to their sources. Unavailable evidence remains missing rather than
being converted to zero and is excluded from aggregation. Recording-quality
checks determine evidence availability and qualify confidence; they are reported
separately rather than treated as an additional cognitive-evidence source.

The participant result page presents an exploratory cognitive monitoring
summary for the current sample, together with its quality limitations. Each
retained feature receives a sample-specific score from 0 to 100, with higher
values indicating more prominent difficulty-oriented patterns from the
AD-related feature inventory rather than better cognitive performance, disease
probability, or diagnosis. \MindBridge{} computes the dimension scores and overall
research-signal index as
registry-priority-weighted means over available feature scores. Core features
receive a weight of 1 and supportive features a weight of 0.5. The resulting
index does not represent general cognitive ability, a diagnosis, or a
probability of dementia. This scope appears alongside data quality, missing
evidence, and the five dimensions so that the result remains tied to the
material supplied in the session. Appendix~\ref{app:score-construction}
specifies the aggregation rule.

To examine how presentation affects interpretation of the same monitoring
evidence, we instantiated three report packages for the user study
(Figure~\ref{fig:participant-interfaces}(d--f)).
All three packages display the same case vignette above the result panel.
In the score-only baseline, the panel displays only the overall 0--100 index. The
five-dimension design removes the aggregate index and asks the reader to
synthesize dimension- and quality-level evidence. The interpretation-first
design retains the index as a secondary reference but places what the sample
shows, what remains unknown, and how quality and missing evidence constrain
interpretation before the number. The score-only package deliberately isolates
an information-minimal numerical presentation, while the other two packages
provide richer forms of explanation. The full study stimuli are specified in
Section~\ref{sec:result-presentation-study}.

These participant-facing alternatives are complemented by a researcher-facing
workbench that exposes the ordered dialogue, input modality, data quality,
missingness, feature-level evidence, and component versions for each session.
Researchers can inspect each observation's source modality and review the
ordered dialogue from the same session.
The researcher-facing report and ordered dialogue views are reproduced in
Appendix~\ref{app:additional-interface-screens}
(Figures~\ref{fig:research-workbench-report}
and~\ref{fig:research-workbench-trace}).

\section{Formative Study of Elicitation and Interpretation}
\label{sec:result-presentation-study}

Ten participants completed an online study with two parts. First, they used
\MindBridge{}'s typed interaction to complete a bounded conversation, immediate
recall after reading, and autobiographical narration (RQ1). Second, they
interpreted reports for two fictional cases, rated and compared three report
packages, and completed a written debrief (RQ2). The fictional reports were
fixed study materials, not outputs generated from participants' own responses.
Table~\ref{tab:human-study-flow} presents the study sequence and the data
collected at each step.

\begin{table*}[htbp]
  \caption{Study procedure and data collected at each step. Case H and Case L
  are fictional cases with relatively high and low sample-level signal indices,
  respectively. Package A, Package B, and Package C present a score-only,
  five-dimension, and interpretation-first report, respectively.}
  \label{tab:human-study-flow}
  \centering
  \small
  \begin{tabular}{@{}p{0.15\linewidth}p{0.34\linewidth}p{0.43\linewidth}@{}}
    \toprule
    Step & Participant activity & Data collected \\
    \midrule
    1. Onboarding & Reviewed the one-sample scope, received a report-package assignment, and reported age band and technology familiarity & Participant-provided ID, assigned package, age band, reading languages, AI-use frequency, and web confidence \\
    2. Bounded chat & Typed an account of the day, answered follow-ups, and completed the chat questionnaire & Ordered dialogue, 14 item-level ratings, a helpful-follow-up example, reported interaction problems, and reported changes to intended wording \\
    3. Complementary tasks & Reconstructed a recently read passage with optional review, then narrated a childhood or another earlier experience & Review count, free and cued recall, prompt choice, narrative fields, eight task-experience ratings, and open feedback \\
    4. First report: Case H & Rated pre-report task significance, viewed the assigned report, explained it in their own words, and completed follow-up items & Meaning, limits, use rationale, quality recall, significance change, confidence, and influential element \\
    5. Second report: Case L & Repeated the report-interpretation task after completing the entire Case H block & The same measures as Case H, treated as second-case responses \\
    6. Interface rating & Rated the report interface after both cases & Six clarity/accessibility items and five capability-specific reliance items \\
    7. Package comparison & Compared all three packages for Case H and ranked them by how well they supported correct understanding & Judgments of interpretability, score fixation, clarity about uncertainty, concern or overreaction, preferred participant report, and first-screen requirements \\
    8. Debrief & Reflected on the chat and report interaction & Five typed responses on first attention, report explanation, quality and uncertainty, follow-up experience, and desired revisions \\
    \bottomrule
  \end{tabular}
\end{table*}

\subsection{Study Design, Participants, and Scope}

Typing deliberately removes microphone, turn-end, ASR, and acoustic variability
from the elicitation study, allowing RQ1 to focus on follow-up relevance and
task-level agency. RQ2 separately examines report comprehension across whole
information packages. Within the formative-study build, GPT-5.4 generates the bounded follow-ups and participants can revise before sending, so the elicitation findings characterize typed interaction with the agent. This study is therefore scoped to the typed configuration, and its fixed fictional reports are not generated from participants' conversations. Evaluation of the speech pathway is left to future work.

We recruited native speakers of Chinese or Japanese through social media.
Participants completed the study online and received compensation at an
effective hourly rate slightly above the applicable local statutory minimum
wage.

We analyzed all ten completed participant records, spanning ages 18--80+ and
varied conversational-AI experience. All participants selected Chinese as a
comfortable reading language; four also selected English and one Japanese.
Table~\ref{tab:human-participants} reports participant-level age,
technology experience, and report assignment. We assigned P1--P10 in session-creation
order; these paper identifiers do not expose participant-provided IDs.

\begin{mbwidetable}
  \caption{Participant context and completed typed-chat records. AI use is
  self-reported frequency; web confidence is rated 1--5. ``Units'' is the
  prototype's language-agnostic word-like-unit count, not a measure of response
  quality. ``First report'' denotes the package assigned for the first fictional
  case, before the later comparison of all three packages. Package A is
  score-only, Package B presents a five-dimension breakdown, and Package C
  presents interpretation before the index.}
  \label{tab:human-participants}
  \centering
  \small
  \begin{tabular}{@{}llllrrr@{}}
    \toprule
    ID & Age band & First report & Conversational-AI use & Web confidence & User turns & Units \\
    \midrule
    P1  & 18--29 & Package B & Almost daily        & 5 & 6 & 97 \\
    P2  & 18--29 & Package C & Several times/week  & 4 & 6 & 141 \\
    P3  & 30--44 & Package B & Several times/month & 4 & 6 & 152 \\
    P4  & 45--59 & Package B & Less than monthly   & 3 & 6 & 122 \\
    P5  & 45--59 & Package C & Never               & 2 & 6 & 142 \\
    P6  & 60--69 & Package C & Several times/month & 4 & 6 & 203 \\
    P7  & 70--79 & Package A & Never               & 2 & 6 & 79 \\
    P8  & 30--44 & Package A & Almost daily        & 5 & 7 & 60 \\
    P9  & 30--44 & Package B & Almost daily        & 5 & 6 & 102 \\
    P10 & 80+    & Package B & Less than monthly   & 3 & 6 & 128 \\
    \bottomrule
  \end{tabular}
\end{mbwidetable}

Participants entered only after online consent, supplied an ID to link study
modules, and could request early deletion of records otherwise retained for six
months on a research-team-restricted server. The questionnaire labels all
reports fictional, frames \MindBridge{} as a research prototype, and states that
one sample cannot characterize overall cognitive state. Quotations below use
P-identifiers.

\subsection{Elicitation Boundary: Follow-Ups and Task-Level Control}

To address RQ1, we used the bounded chat and two complementary tasks to
examine changes in participants' contributions and their control over
their responses.

The study-specific chat instantiates the bounded policy in
Section~\ref{sec:mindbridge}. It instructs the model to ask one follow-up at a
time about a detail from the participant's account of that day. Participants
can revise their responses, choose what to disclose, skip, or finish.
Role-separated turns and visible system states preserve the interaction
sequence. Appendix~\ref{app:additional-interface-screens} reproduces the typed
interface used to isolate this interaction from speech capture
(Figure~\ref{fig:typed-study-interface}).

We assessed both the elicitation benefit and interactional cost of these
follow-ups. Fourteen study-specific five-point items cover legibility, latency,
relevance, one-question clarity, additional detail, control, repair, evaluative
concern, effort, and willingness to repeat. An open item elicits a helpful
follow-up example, while two checklists capture interaction problems and
whether participants delete, shorten, reorganize, add, or rephrase material.
These are formative items rather than a validated scale or a comparison with a
non-adaptive interviewer.

The same participants completed the two complementary typed tasks described in
Section~\ref{sec:mindbridge}. Immediate recall hides a fixed passage before
response and records optional five-second reviews alongside the reconstruction,
detail responses, ratings, and feedback. Autobiographical narration records
whether participants use the childhood prompt, redirect it, or skip, together
with their narrative fields, ratings, and feedback. We analyzed both modules as
probes of task interaction and evidence provenance, not memory accuracy or
person-level ability.

\subsection{Interpretation Boundary: Report Hierarchies}

Participants viewed two fictional cases with relatively high (Case H) and low
(Case L) sample-level signal indices. Each case was presented through one of
three report packages: Package A, Package B, or Package C.
Table~\ref{tab:report-cases} summarizes the case settings and the interpretation
questions they were designed to probe.

\begin{table*}[t]
  \caption{Two fictional cases used in the report-interpretation study. Case H and Case L
  refer to relatively high and low sample-level signal indices, not levels of
  overall cognitive ability.}
  \label{tab:report-cases}
  \centering
  \small
  \begin{tabular}{@{}p{0.10\linewidth}p{0.43\linewidth}p{0.39\linewidth}@{}}
    \toprule
    Case & Case setting & Interpretation question \\
    \midrule
    Case H & Relatively high signal index; poor sleep, television noise, fair data quality, and missing speech evidence & Whether readers treat the high index as conclusive despite limited data quality and missing evidence. \\
    Case L & Relatively low signal index; good data quality, but only a familiar-routine account with no independent memory, attention, or reasoning task & Whether readers infer good overall cognitive functioning from a low index in a narrowly sampled task. \\
    \bottomrule
  \end{tabular}
\end{table*}

A prominent number may anchor interpretation before readers inspect its source,
quality, or limitations. We therefore compared three whole presentation
packages built from the same evidence within each case. Beneath the shared
vignette, Package A displays only the aggregate. Package B foregrounds quality,
dimensions, and expandable evidence. Package C places sample scope and limits
before a secondary index. Because information,
hierarchy, aggregation, disclosure, and styling vary together, responses cannot
be attributed to score placement alone. Figure~\ref{fig:participant-interfaces}(d--f)
shows the three default screens for Case~H side by side.

Each case first elicited an open explanation to a fictional family member, coded
as sample-scoped or as converting the value into overall cognitive ability or a
clinical conclusion. Later items examine numerical meaning, quality or
uncertainty recall, evidence-consistent use, clarity, accessibility, and
capability-specific reliance. Perceived evidential significance was measured
before the report and after the complete case block, so its change cannot be
attributed to the report page alone. Participants then compared and ranked
Package A, Package B, and Package C and completed an open debrief; we did not
collapse these outcomes into a generic trust or usability score.

\subsection{Study Procedure and Analysis}

For the report-interpretation phase, the instrument assigned Package A,
Package B, and Package C to two, five, and three participants for Case H.
Nine participants retained their initial package for Case L; a later instrument
configuration assigned P9 to Package B for Case H and Package C for Case L,
yielding two, four, and four participants for Case L. These
unblocked assignments and the configuration difference provide varied exposures
rather than an interface-effect comparison. A landing page first stated the
one-sample scope and the roles of quality, missingness, and context.
For each case, an open explanation preceded quality-recall and other closed
items. Case L followed Case H in fixed order and was treated as a second-case response.
Participants then rated interface capabilities, compared all three packages,
and completed a typed debrief. This comparison supports formative critique, not
a causal within-participant test; no separate spoken-interview or
facilitator-note corpus was collected.

The system does not record dwell time, disclosure expansion, scrolling, or
inspection order; first attention is self-report rather than a gaze or log
measure.

For analysis, we report five-point items as median [range], categorical
responses as $n/N$, and perceived-significance change as each participant's
post-minus-pre value.
For the complementary tasks, we report review and prompt-choice actions, the
same item-level summaries, and explicit recurring concerns in open feedback.
We did not derive a recall score, compare task performance by age, or interpret
response content as cognitive status.
All ten included records enter every denominator; no response was imputed. Given
the small, uneven first-exposure groups and whole-package manipulation, we used
no significance tests, confidence intervals, age comparisons, or
interface-superiority claims.

We applied a transparent, a priori descriptive rule to the two open report
explanations. A response is \emph{sample-scoped} if it explicitly limits the
page to the presented sample or task, or rejects an inference about stable,
long-term, or overall cognitive ability. It is \emph{overgeneralized} if it
treats the displayed value as a measure of the fictional person's cognitive
level or as a clinical conclusion. Quality recall is counted when the response
names data quality, missing evidence, or an unmeasured task. The closed
evidence-consistent-use items are reported separately so that selecting the
prompted answer is not equated with producing a calibrated open explanation.
These simple codes were not independently double-coded; we therefore provide
counts, participant identifiers, and contrasting excerpts rather than a
reliability statistic.

For the chat transcripts, helpful-follow-up responses, task feedback,
checklists, and debrief, we used descriptive content analysis to identify
interaction mechanisms and design requirements. We checked each reported
unsupported assumption against the stored dialogue and retained negative cases.
The corpus comprises
participant-entered Chinese text only; concise excerpts below are English
translations. We make no claim of thematic saturation, and quotation is used to
make a mechanism concrete rather than to imply prevalence.

Accordingly, the study supports mechanism-level rather than causal or
population-level claims. Case L is a second-case response, with same-hierarchy
transfer for nine participants, and the later comparison among Package A,
Package B, and Package C supports whole-package critique rather than isolated
interface effects. Full
limitations appear in
Appendix~\ref{app:detailed-limitations}.

\section{Results}
\label{sec:human-study-results}

In this section, we address RQ1 by examining chat transcripts, questionnaire
responses, and records of passage views and prompt choices. For RQ2, we analyze
report explanations, ratings of willingness to rely on the reports, and
comparisons and rankings of the three report packages. For RQ3, we examine
\MindBridge{}'s analysis pipeline using fixed recordings and controlled synthetic
checks.

\begin{figure*}[t]
  \centering
  \includegraphics[width=\textwidth]{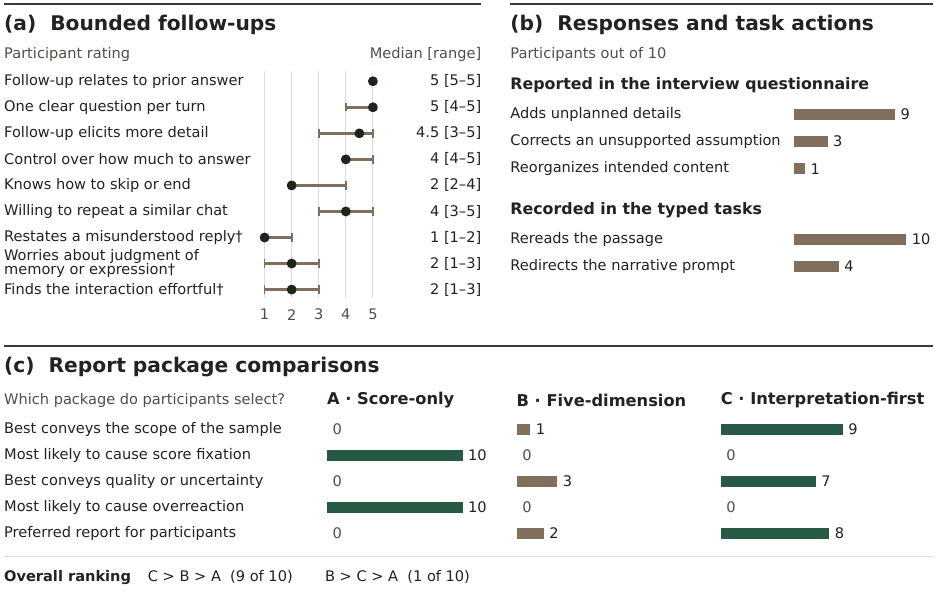}
  \caption{Descriptive results from the formative study at the elicitation and
  interpretation boundaries ($N=10$). (a): Responses to study-specific
  five-point items about the bounded follow-up interaction. Points indicate
  medians, horizontal lines indicate observed ranges, and $\dagger$ marks
  negatively worded interaction-cost items, for which higher values indicate a
  less favorable experience. (b): Counts from the interview questionnaire
  cover changes to intended responses and assumptions participants noticed in
  follow-ups. The task logs record passage views and prompt choices.
  (c): Participant counts from the
  post-exposure comparison of three whole report packages: Package A
  (score-only), Package B (five-dimension), and Package C
  (interpretation-first). Green bars mark the most frequently selected package
  for each question. The last line ranks packages by how well participants
  thought they supported correct understanding.
  Panels (b) and (c) report counts out of 10. These descriptive
  findings characterize this formative sample but do not estimate causal
  interface effects or establish package superiority.}
  \Description{Three panels summarize the formative study with ten
  participants. Panel (a) shows medians and observed ranges for nine ratings of
  bounded follow-ups. Relevance and one-question clarity have medians of five,
  additional detail has a median of 4.5, response-amount control and willingness
  to repeat have medians of four, skip-or-end discoverability has a median of
  two, and the three negatively worded cost items have medians of one or two.
  In panel (b), nine participants reported adding unplanned details, three
  reported a follow-up with an unsupported assumption, and one reported
  reorganizing intended content. All ten reopened the passage for review and
  four redirected the autobiographical prompt. In panel (c), all ten chose
  Package A on the score-fixation item and on the concern-or-overreaction item.
  Nine chose Package C for conveying sample scope, seven for communicating
  quality or uncertainty, and eight for participant use. When ranking the
  packages by support for correct understanding, nine ranked Package C first
  and one ranked Package B first. All ten ranked Package A last.}
  \label{fig:human-results-overview}
\end{figure*}

Participants reported adding detail in response to follow-ups and identified
questions that introduced unsupported assumptions. Most participants bounded
report interpretations to the current sample, while contrasting cases revealed persistent
overgeneralization across report packages. Package comparisons favored
explanations of scope and limitations before aggregate scores.
Figure~\ref{fig:human-results-overview} summarizes selected questionnaire
responses and task-action counts.

\subsection{RQ1: Follow-Ups and Task Controls Shape the Record}


The structured responses show why ``relevant'' and ``neutral'' are different
interaction properties. All participants gave the maximum relevance rating,
and nine reported adding details they had not initially planned to include; P9
instead shortened the intended response while still identifying a meal-detail
prompt as helpful. Helpful prompts selected a participant-introduced anchor and
requested one level of specification. For example, P1 reported that a chart-related
question elicited the experiments, chart form, colors, and axis labels.


Three participants reported a follow-up with an unsupported assumption.
P2 explicitly corrected the interviewer during the conversation. Asked whether
they next revised the methods or the wording, P2 explained that they prepared
for a class. P1 and P6 continued their accounts and identified the assumption
in the debrief. P1 noted that ``returned'' implied they had left
the laboratory. P6 noted that ``unexpected delight'' assigned a feeling they
had never expressed.

Perceived agency was also uneven across levels of the interface. Participants
agreed that they controlled response length, rarely reported rewriting, and
generally found the task low effort. Yet the median for knowing how to skip a
question or end the task was only 2. Control exists in the interaction policy
but is not sufficiently discoverable in the interface. Together these results
motivate grounded, open continuation questions and persistent skip and finish
affordances rather than relying on participants to express refusal inside the
conversation.

The complementary tasks expose agency at a different granularity. All ten used
optional review in immediate recall, with a median count of 2 (range 1--4).
Task clarity was 4.5 [4--5], support for expressing remembered content was 4
[3--5], review-control clarity was 4 [2--5], and effort was 3 [2--5]. Five
described the fixed five-second window as too short or rushed. Retaining review
count thus distinguishes responses produced under different stimulus exposure.

All ten completed autobiographical narration: six used the childhood prompt, four
redirected it, and none skipped. Expression was 4 [3--5], perceived control over
disclosure and skipping was 5 [4--5], effort was 3 [2--5], and discomfort was 1
[1--2]. The task asked for the story, its context, and a memorable detail.
Seven participants reported overlap among these fields.

The debrief revealed problems with follow-ups even when participants answered
without challenging the question. Participants used passage review and prompt redirection
to control their task responses, although some were unsure how to skip or
end the interview.

\subsection{RQ2: Report Comprehension and Capability-Specific Reliance}

Table~\ref{tab:report-results} summarizes how participants interpreted each
case and which report capabilities they would rely on.
The shared vignette also supplied cues about data quality or task scope that
participants could draw on when answering the recall question.
Figure~\ref{fig:human-results-overview}(c) shows their judgments after they
compared all three packages for Case~H.

\begin{table*}[t]
  \caption{Report interpretation and capability-specific reliance ($N=10$).
  Case H is the high-index case shown first; Case L is the low-index case shown
  after the complete Case H block. Five-point items and post-minus-pre
  significance changes are median [range].}
  \label{tab:report-results}
  \centering
  \small
  \begin{tabular}{@{}p{0.72\linewidth}r@{}}
    \toprule
    Outcome & Observed value \\
    \midrule
    Sample-scoped open interpretation, Case H & 8/10 \\
    Overgeneralizes to person-level cognitive ability, Case H & 2/10 \\
    Recalls quality, missingness, or uncertainty, Case H & 9/10 \\
    Selects evidence-consistent use, Case H & 10/10 \\
    Perceived-significance change, Case H & +1 [0--2] \\
    Interpretation confidence, Case H & 4 [3--5] \\
    Sample-scoped open interpretation, Case L (second case) & 8/10 \\
    Overgeneralizes to person-level cognitive ability, Case L & 2/10 \\
    Recalls quality, missingness, or unmeasured tasks, Case L & 8/10 \\
    Selects evidence-consistent use, Case L (second case) & 8/10 \\
    Perceived-significance change, Case L & +1 [0--2] \\
    Interpretation confidence, Case L & 4.5 [3--5] \\
    \midrule
    Willingness to rely on summarizing one sample & 4 [2--5] \\
    Willingness to rely on flagging missing or low-quality data & 5 [1--5] \\
    Willingness to rely on comparing repeated samples & 4 [2--5] \\
    Willingness to rely on inferring information absent from page & 1 [1--5] \\
    Willingness to rely on an overall score without evidence & 1 [1--5] \\
    \bottomrule
  \end{tabular}
\end{table*}

Eight participants limited their interpretations to the current sample in
both the high-index case (Case H) and the low-index case (Case L). The two
participants initially assigned the score-only report (Package A) provide
a useful contrast. P7 said of Package A,
``I can only see 69 and do not know what the score is made of,'' then refused to
extend it beyond the noisy sample. P8 instead described 69 as a cognitive level
that was ``passing but not particularly good,'' and later interpreted 23 as
cognitive decline that warranted a hospital visit. P8 reported maximum
willingness to rely on an overall score without checking its evidence and was
highly confident in the explanation of Case L. P7 reported the minimum willingness
to rely on that score and lower confidence. Thus, the bare number does not
determine misunderstanding; it creates an interpretive vacuum in which prior
score conventions can dominate.

P9 supplies a complementary negative case under richer hierarchies. After
viewing the five-dimension report (Package B) for the high-index case (Case H),
P9 acknowledged noise and missing speech evidence but still described the page
as the person's ``cognitive testing level.'' After
viewing the interpretation-first report (Package C) for the low-index case
(Case L), P9 interpreted the low value as a cognitive problem requiring medical
attention, despite the preceding scope and missing-task explanation. P9 also
reported maximum willingness to infer absent information and rely on an overall
score without its evidence. Explanation order can therefore support a bounded
reading without guaranteeing it; prior expectations remain part of the
interpretation boundary.

The ordering of measures exposes a second distinction. P8 and P9 both selected
the evidence-consistent closed answer for Case H after producing person-level open
interpretations, making the prompted Case H result 10/10 even though only eight open
explanations were sample-scoped. Both selected the overgeneralized option for Case L.
The other eight participants selected the bounded option and named the missing
speech evidence for Case H and the unmeasured memory, attention, or reasoning tasks
for Case L. Their suggested action was not to convert a single result into a decision,
but to record context, recollect under better conditions, and compare multiple
samples or tasks.


At the group level, willingness to rely on the report varied by capability.
Median ratings were higher for summarizing one sample (4 [2--5]), flagging
missing or low-quality data (5 [1--5]), and comparing repeated samples
(4 [2--5]) than for inferring information absent from the page or relying on
an overall score without evidence (both 1 [1--5]). Ratings nevertheless varied
widely across participants. Perceived evidential significance increased by a
median of 1 [0--2] in each case. Because the post-rating followed the complete
case block, this change reflects responses to the full interpretation activity
and cannot be attributed to the report page alone.

\subsection{Cross-Package Critique: Explanation Before Aggregation}


Self-reported first attention helps explain the comparisons. Both readers of
the score-only report (Package A) first noticed the large number, all three
readers of the interpretation-first report (Package C) first noticed the
preceding ``this sample shows'' or ``cannot show'' blocks, and all five readers
of the five-dimension report (Package B) entered through a numerical profile.
Although not gaze traces, these reports show how hierarchy supplies different
entry points into the same evidence.

After seeing all packages, all ten chose the score-only report (Package A)
when asked which was most likely to draw attention to the number alone. All ten
also chose it when asked which caused the most concern or seemed most likely to
prompt an overreaction. P2 likened its 69/100 to an examination grade. Nine
ranked the interpretation-first report (Package C) highest for supporting
correct understanding. Eight preferred it for participant use because scope,
limitations, and use precede the index. P8 and P10 instead valued the inspectable
dimensions and evidence in the five-dimension report (Package B), and three
selected Package B as clearest for quality and uncertainty.

Most participants interpreted the reports as descriptions of the current
sample, and reliance ratings were higher for summarizing observed evidence
than for unsupported inferences. Participants favored explanations before
scores, although some still drew conclusions about overall cognitive ability
after viewing the more detailed reports.

\subsection{Connecting the Two Boundaries into Design Requirements}

The first-screen checklist distinguishes information that participants wanted
immediately from evidence they were willing to inspect on demand. Nine selected a
one-sample statement, the aggregate value's name and scope, and data quality;
eight selected a plain-language interpretation and missing or unusable evidence.
Seven selected the warning that low-looking results cannot fill in unmeasured
cognitive abilities, and four required all five dimensions and their sources on
the first screen. This pattern favors progressive disclosure: the boundary and
quality remain proximal to any summary, while detailed provenance stays
reachable without dominating the initial reading.

\begin{table*}[t]
  \caption{Design implications for \MindBridge{} informed by the formative study.}
  \label{tab:human-design-requirements}
  \centering
  \small
  \begin{tabular}{@{}p{0.26\linewidth}p{0.70\linewidth}@{}}
    \toprule
    Design focus & Design implication \\
    \midrule
    Follow-up wording & Ground follow-ups in stated details, avoid unsupported assumptions, and retain links to source turns. \\
    Skip and finish controls & Keep labeled Skip and Finish controls visible beside the response field. \\
    Review and prompt choice & Record review and prompt choices, allow adjustable review timing, and avoid redundant questions. \\
    Score presentation & Explain sample scope and limitations before the index and keep them visible alongside it. \\
    Evidence detail & Provide a concise initial explanation with expandable dimensions and supporting evidence. \\
    Missing evidence & Distinguish unavailable evidence from low values and identify abilities or modalities not assessed. \\
    Readability and navigation & Support adjustable text size, sufficiently large controls, and clearly labeled access to evidence. \\
    \bottomrule
  \end{tabular}
\end{table*}

Table~\ref{tab:human-design-requirements} summarizes the design implications.
Across both boundaries, the interface must reveal how prompts, review, and
participant choices shape an answer, and how aggregation shapes its
interpretation. Human-centered explainability therefore requires both
researcher-facing records and participant-facing controls, with limitations
proximal to values and an information order that favors a bounded reading.

\subsection{RQ3: Technical Evaluation of the Multimodal Analysis Pipeline}
\label{sec:technical-boundary-checks}


To address RQ3, we conducted three complementary checks of \MindBridge{}'s
analysis pipeline. First, we compared model configurations on fixed
recordings, examining overall accuracy and errors for each corpus class.
Second, we tested whether the pipeline distinguished predefined levels of
communication difficulty across text and synthetic speech. Third, we
examined error patterns across translated and synthesized language versions
of the same cases. These evaluations characterize pipeline behavior; they
do not validate the participant-facing five-dimension summary or
research-signal index as measures of cognitive status.
Appendix~\ref{app:technical-details} provides the fixed-recording manifest,
construction details, manipulation statistics, and cross-lingual protocol.

\paragraph{Fixed-recording modality audit.}
We analyzed 109 English Cookie Theft sessions from DementiaBank
\cite{becker1994natural}: 48 Control recordings, denoted Corpus-C, and 61
recordings labeled ProbableAD, PossibleAD, or MCI, combined as Corpus-NC for
this binary audit. Corpus-NC serves as the positive class; predicting it for
every recording would yield 56.0\% accuracy. Text models received participant
turns only, whereas speech and speech-plus-text models could receive the complete
recording, including investigator speech annotated in 106 sessions. Input
scope is therefore entangled with modality.
Table~\ref{tab:model-modality-comparison} reports configuration-level zero-shot
accuracy on this fixed manifest.

\MindBridge{} uses GPT-5.4 for transcript analysis and structured evidence
fusion, and gpt-realtime-2 for speech interaction and qualitative speech
analysis. In the fixed-recording evaluation, these models achieved the
highest accuracy among the tested text-only (70.3\%) and speech-only
(66.1\%) configurations, respectively. These results characterize the
individual evaluation configurations, not their combined deployment.

Aggregate accuracy alone, however, obscures important failure modes. Under a
common open-model protocol, Qwen2-Audio and Ultravox predicted
Corpus-NC for every recording, Step-Audio did so for 108 of 109, and
Qwen2.5-Omni reached 62.4\% accuracy with 78.7\% sensitivity and 41.7\%
specificity. This is a
failure-mode audit, not a controlled modality ablation; it shows that an
aggregate rank can conceal class collapse and interviewer exposure.

\begin{table}[t]
  \caption{Accuracy of single-modality and multimodal configurations on the
  fixed 109-recording manifest. $\dagger$ denotes model components used in
  \MindBridge{}. GPT-5.4 also performs evidence fusion. Reported accuracies describe
  the evaluated configurations, not their combined deployment.}
  \label{tab:model-modality-comparison}
  \centering
  \small
  \begin{tabular}{@{}llr@{}}
    \toprule
    Model & Modality & Accuracy (\%) \\
    \midrule
    \multicolumn{3}{@{}l}{Single modality} \\
    GPT-5.4$^{\dagger}$ & Text         & 70.3 \\
    Qwen2.5-Omni-7B  & Text         & 60.6 \\
    Qwen3-Omni-30B   & Text         & 65.1 \\
    \addlinespace[2pt]
    gpt-realtime-2$^{\dagger}$ & Speech        & 66.1 \\
    Audio-Flamingo-3 & Speech     & 44.0 \\
    Qwen2.5-Omni-7B  & Speech        & 55.0 \\
    Qwen3-Omni-30B   & Speech        & 56.9 \\
    \midrule
    \multicolumn{3}{@{}l}{Multimodal} \\
    gpt-realtime-2   & Speech + text & 58.7 \\
    \addlinespace[2pt]
    Audio-Flamingo-3 & Speech + text & 54.6 \\
    Qwen2-Audio-7B-Instruct  & Speech + text & 56.0 \\
    Qwen2.5-Omni-7B         & Speech + text & 62.4 \\
    Step-Audio-2-mini        & Speech + text & 56.9 \\
    Ultravox-v0.4.1-12B     & Speech + text & 56.0 \\
    Qwen3-Omni-30B   & Speech + text & 59.6 \\
    \bottomrule
  \end{tabular}
\end{table}

\paragraph{Controlled synthetic-speech probe.}
We held a six-to-nine-event graph fixed for each of 20 neutral older-adult
personas and rendered it at four programmed communication-difficulty levels as
text and synthetic speech. All eight transcript- and waveform-derived
manipulation measures changed in the programmed direction ($p<.001$). A blinded
Qwen2.5-Omni probe completed 240 inferences across text, speech, and their
combination. Table~\ref{tab:synthetic-qwen-ordinal} reports the resulting
ordinal-sensitivity measures. Although every mode retained a non-decreasing
order for 19 of 20 personas, exact level recovery fell from 85.0\% for text to
26.25\% for speech and 50.0\% for speech plus text.
These findings concern discrimination among programmed synthetic conditions,
rather than cognitive-status assessment in naturally produced human speech.

\begin{table*}[t]
  \caption{Blinded ordinal-sensitivity results for Qwen2.5-Omni-7B. Condition
  recovery is agreement with the programmed synthetic level, not accuracy on
  naturally produced human samples. Strict-order measures treat ties as
  failures. QWK denotes quadratic weighted $\kappa$.}
  \label{tab:synthetic-qwen-ordinal}
  \centering
  \setlength{\tabcolsep}{3.3pt}
  \begin{tabular}{@{}lrrrrrrrrr@{}}
    \toprule
    Mode & Valid & Recovery & MAE & Pooled $\rho$ & Within-persona $\rho$ &
    QWK & Adjacent strict & All-pairs strict & Full strict \\
    \midrule
    Text        & 80/80 & \textbf{85.00\%} & \textbf{.150} & \textbf{.940} &
    \textbf{.959} & \textbf{.939} & \textbf{83.33\%} & \textbf{90.83\%} & 50.00\% \\
    Speech       & 80/80 & 26.25\% & .975 & .764 & .867 & .455 & 46.67\% & 62.50\% & 0.00\% \\
    Speech+text  & 80/80 & 50.00\% & .550 & .837 & .895 & .691 & 58.33\% & 75.00\% & 0.00\% \\
    \bottomrule
  \end{tabular}
\end{table*}

\paragraph{Cross-lingual stress test.}
We also paired the 109 English cases with Chinese and Japanese translations and
synthetic speech that preserve annotated disfluencies. Speech-plus-text
accuracy was identical for Chinese and Japanese (62.39\%), yet
sensitivity/specificity shifted from 52.46\%/75.00\% to 86.89\%/31.25\%.
Translation, synthesis, and language changed together, so this result cannot
isolate a language effect or support a multilingual usability claim.


Together, the three checks show that \MindBridge{}'s analysis pipeline cannot
be characterized by overall accuracy alone. Some model configurations
assigned all recordings to the same class. In the synthetic probe,
predefined difficulty levels were recovered less accurately from speech
than from text. The translated and synthesized language versions also
yielded identical overall accuracy but different error patterns.

\FloatBarrier

\section{Discussion}

\subsection{RQ1: The Adaptive Interviewer Is Part of the Measurement}

%
%

The RQ1 findings suggest that an adaptive interviewer should be evaluated
not only by how relevant its questions seem, but also by how they shape
the resulting evidence.
Grounding a follow-up in a participant's answer does not remove the need to
check for assumptions introduced by the question.

Follow-ups should retain links to participant-provided details, while Skip and
Finish controls should be visible rather than dependent on a participant
knowing how to refuse in conversation. Rereading and prompt redirection should
likewise be recorded as task conditions because they change stimulus exposure
and the experience narrated. Retaining this context helps researchers interpret
a response in relation to how it was elicited, rather than attribute every
feature to the participant.

\subsection{RQ2: The Result Interface Is Part of the Inference}

RQ2 supports explanation before aggregation as a design direction, not a
universally superior interface. Reports should state sample scope, data quality,
and missing evidence before the index, with dimensions and source evidence
available on demand. This separates the information needed for an initial
interpretation from the detail needed to inspect its basis.

Clear presentation does not guarantee a bounded interpretation. Score direction
needs explanation beside each value: a label such as ``Discourse fluency'' can
suggest stronger ability even when higher values indicate more prominent
difficulty-related patterns. Evaluations should therefore examine readers' own
explanations and intended uses, rather than infer understanding from report
preference, confidence, or prompted answers alone.

\subsection{RQ3: The Analysis Pipeline Is Part of the Evidence Chain}

RQ3 suggests that an output should be interpreted alongside the conditions
under which it was produced. The fixed-recording comparison does not isolate
modality effects because model configurations and input scope varied together.
Translation and speech synthesis also prevent isolating language effects.

For researchers, a workbench should pair outputs with the model configuration,
input source, modality, language, and missing evidence, and provide access to
errors for each dataset class when inspecting evaluations. For participants,
reports should identify the evidence actually used without implying that all
modalities contributed equally. This separation allows detailed investigation
of pipeline failures without requiring participants to interpret benchmark
diagnostics.

\subsection{Human-Centered Explainability Across the Evidence Process}

Across RQ1--RQ3, explainability concerns how cognitive-monitoring evidence is
obtained, analyzed, and presented (Figure~\ref{fig:monitoring-workflow}).
A feature-level explanation cannot substitute for a record of prompts and task
conditions, while a complete dialogue log cannot ensure that a report
communicates the limits of its claims.

For \MindBridge{}, the design goal is to link each report claim to its supporting
observations, the inputs used to derive them, and the task conditions under
which those inputs were collected. This makes participant control and
researcher inspection complementary: one enables choices during collection;
the other exposes how those choices and subsequent analysis shaped the result.
The same principle may inform adaptive interviews, health-intake agents, and
educational or reflective systems that both elicit and evaluate responses.
Traceability enables scrutiny, but does not itself establish that a judgment
is valid.

\subsection{Scope of Claims and Next Steps}

%

These findings inform the design of cognitive monitoring rather than establish
clinical validity. Clinical input came from one psychiatry resident, and the
small, mixed-age user study examined typed interaction rather than speech use
or memory performance. Whole-package comparisons do not establish causal
interface effects or age differences. The technical checks are limited by
corpus composition, differing input scopes, and translated or synthesized
speech. Appendix~\ref{app:detailed-limitations} details these constraints.

Next evaluations should involve a broader target-user sample using speech,
examining microphone-state clarity, latency, turn-ending controls, and
transcription repair. Behavioral records should be paired with open
explanations to assess whether participants can control collection and
understand what their results support. Independent characterization of the
speech evidence is also needed.

\section{Conclusion}

\MindBridge{} is an interactive, multilingual, and multimodal cognitive monitoring
tool that connects two human-facing processes through multimodal analysis: an
interviewer helps form the evidence, the analysis can preserve or compress it,
and a result interface helps form its interpretation. The formative study turns
these processes into concrete requirements: ground follow-ups in
participant-provided anchors, make Skip and Finish controls persistent,
preserve follow-ups, stimulus review, and prompt choice in their corresponding
records,
and place sample scope, quality, and missingness before and beside any
aggregate. Fixed-recording and synthetic audits show why modality, language,
and class-conditional behavior must remain inspectable. A broader study of
naturally spoken interaction is therefore the next required test of this
interactive and explainable cognitive monitoring approach.

\bibliographystyle{ACM-Reference-Format}
\bibliography{sample-base}

\clearpage
\appendix

\FloatBarrier

\section{Additional System Interface Screens}
\label{app:additional-interface-screens}

The following screens document how the interaction and provenance mechanisms
described in the main text appear at full-page scale. The participant-facing
set contains the typed bounded-follow-up interface used in the formative study
and the two complementary task interfaces. Together, these screens expose
where participants revise or submit a response, review the reading stimulus,
redirect an autobiographical prompt, or skip a task. They therefore make the
controls that shape the elicited record visible rather than presenting the
tasks only as procedural descriptions.

The researcher-facing set uses one synthetic demonstration session to show the
other side of the evidence chain. The report identifies the source modality,
quality, missing evidence, and potential confounders. The dialogue view shows
the turns from that session in order, with the speaker role and input mode.
These screens illustrate the intended connection between participant-facing
control and researcher-facing traceability; they are interface documentation,
not additional study outcomes. All displayed IDs, dialogue, and results are
fictional or synthetic and do not show participant data.

\begin{figure*}[p]
  \centering
  \includegraphics[width=\textwidth,height=0.82\textheight,keepaspectratio]{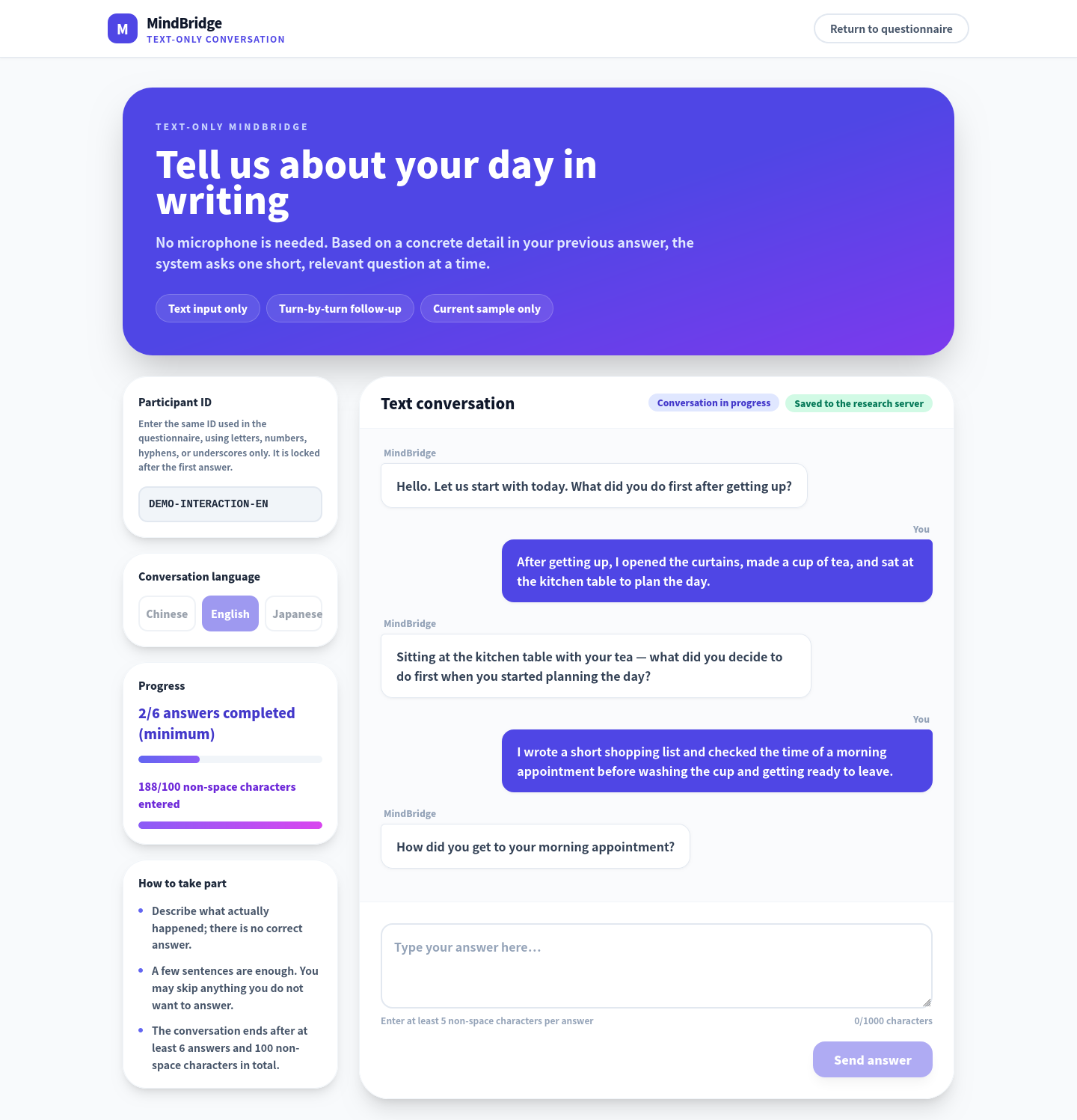}
  \caption{Study-specific typed bounded-follow-up interface. The left column
  exposes participant context, language, progress,
  and interaction guidance; the conversation panel retains ordered turns and
  gives the participant control over revision and submission.}
  \Description{A MindBridge typed-conversation screen shows a fictional
  participant ID, language controls, progress indicators, interaction guidance,
  four alternating dialogue bubbles, and a typed-answer composer.}
  \label{fig:typed-study-interface}
\end{figure*}

\begin{figure*}[p]
  \centering
  \includegraphics[width=\textwidth,height=0.82\textheight,keepaspectratio]{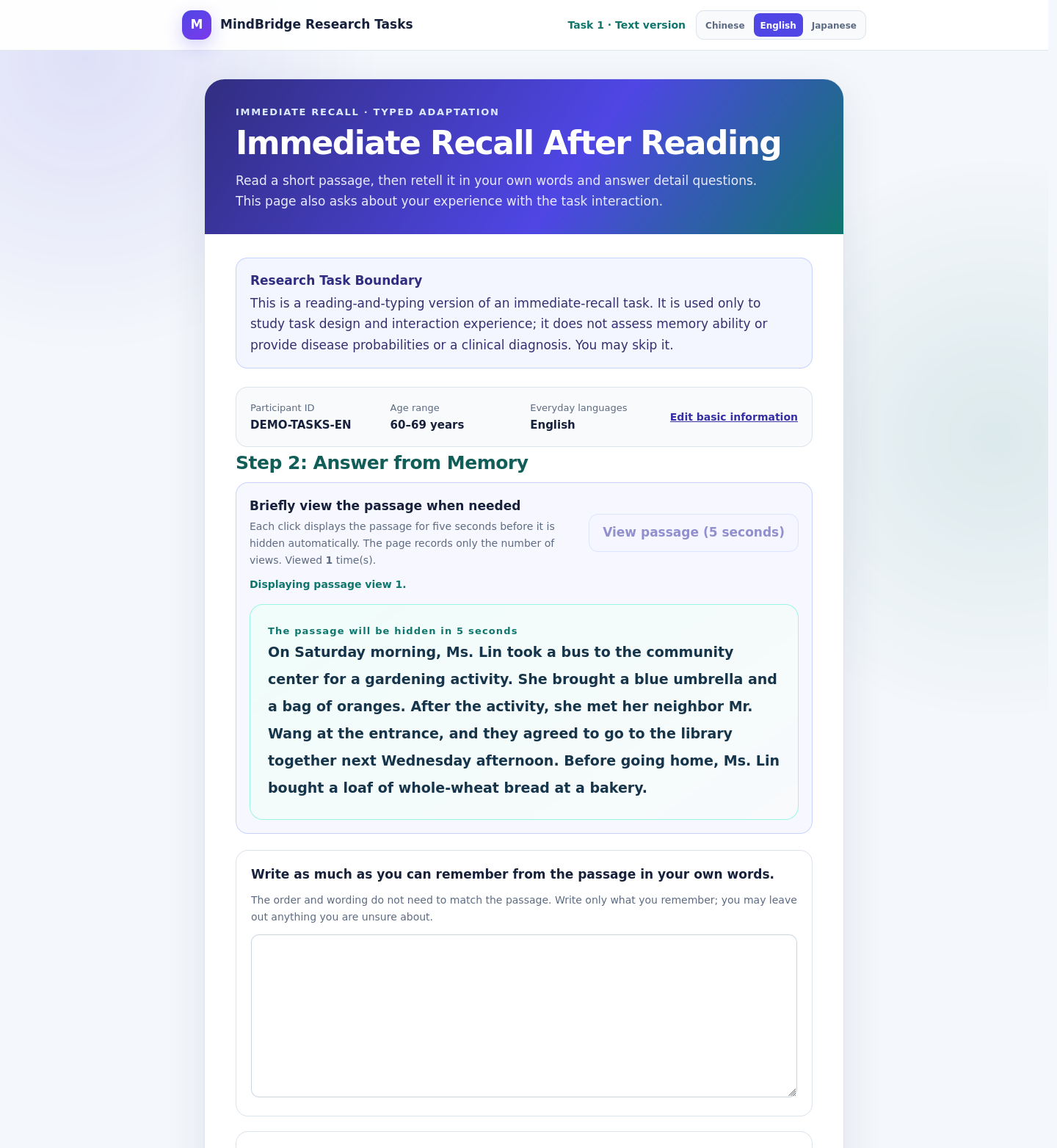}
  \caption{Immediate-recall-after-reading task. The interface makes the
  short source passage, optional five-second review, participant context, and
  free-recall response area visible together.}
  \Description{A MindBridge research-task screen shows the Immediate Recall
  After Reading title, a research-boundary notice, fictional participant
  context, a passage with a five-second review control, and an empty recall
  response box.}
  \label{fig:immediate-recall-screen}
\end{figure*}

\begin{figure*}[p]
  \centering
  \includegraphics[width=\textwidth,height=0.82\textheight,keepaspectratio]{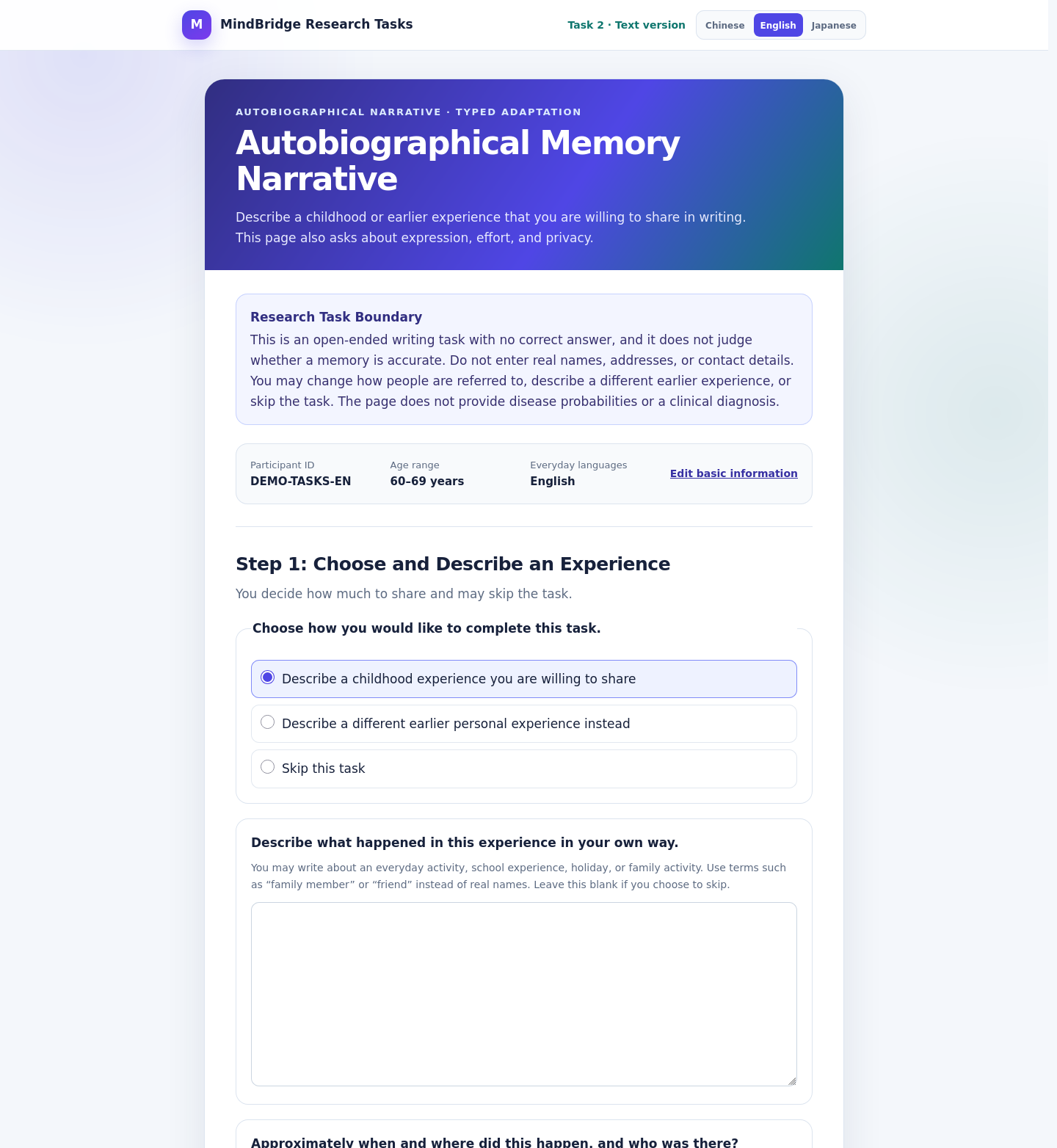}
  \caption{Autobiographical-memory-narration task. Participants choose
  a childhood experience, another earlier personal experience, or an explicit
  skip option before deciding what to disclose in writing.}
  \Description{A MindBridge research-task screen shows the Autobiographical
  Memory Narrative title, a research-boundary notice, fictional participant
  context, three experience-choice controls including Skip this task, and a
  narrative response area.}
  \label{fig:autobiographical-memory-screen}
\end{figure*}

\begin{figure*}[p]
  \centering
  \includegraphics[width=\textwidth,height=0.82\textheight,keepaspectratio]{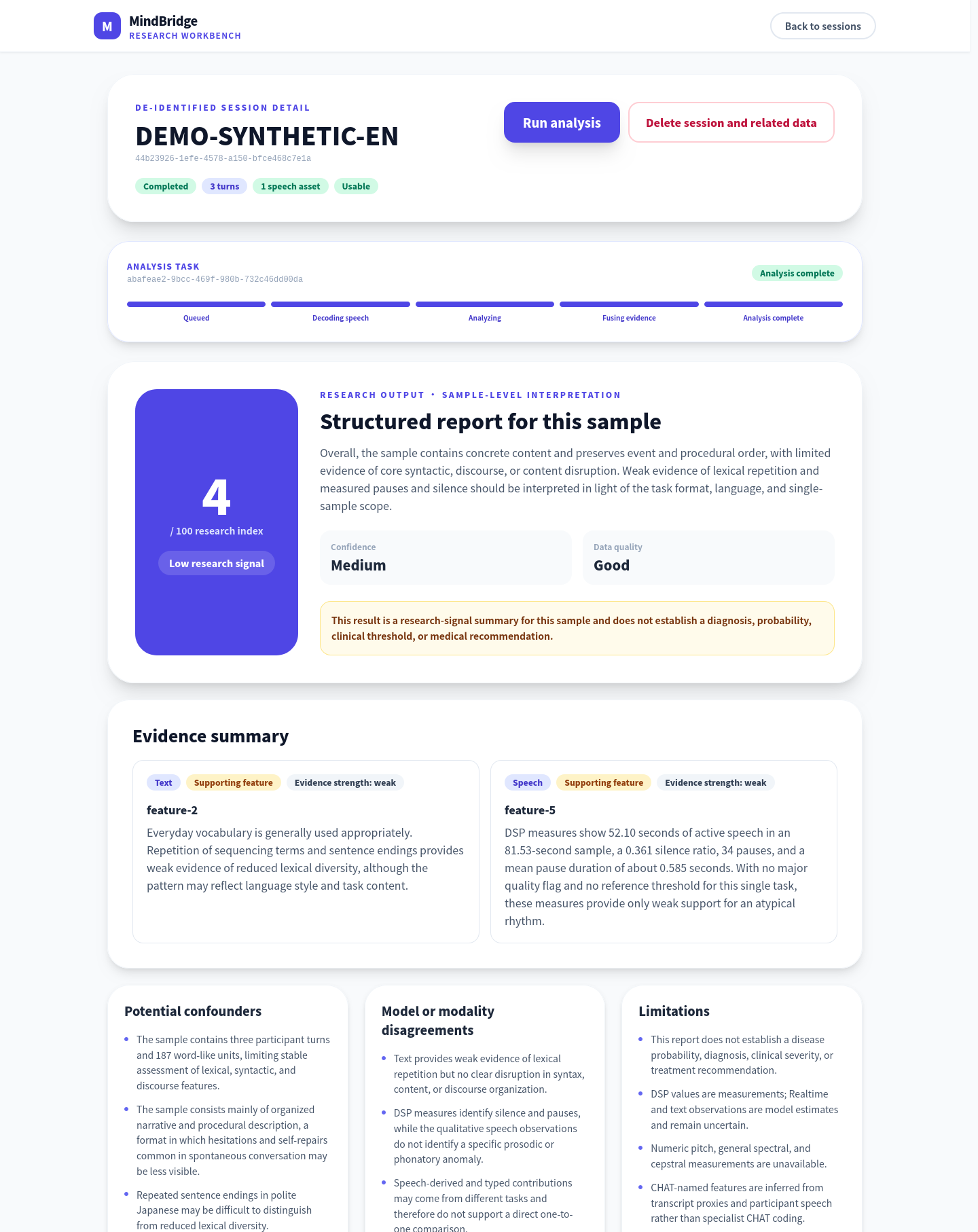}
  \caption{Researcher-facing report view for a synthetic demonstration session.
  The English translation presents the interpretation of the Japanese sample,
  data quality, observations labeled by source modality, potential confounders,
  modality disagreements, and limitations.}
  \Description{A MindBridge research-workbench screen for a synthetic session
  shows analysis status, a structured report with a secondary research index,
  confidence and data-quality cards, evidence summaries, potential confounders,
  modality disagreements, and limitations.}
  \label{fig:research-workbench-report}
\end{figure*}

\begin{figure*}[p]
  \centering
  \includegraphics[width=\textwidth]{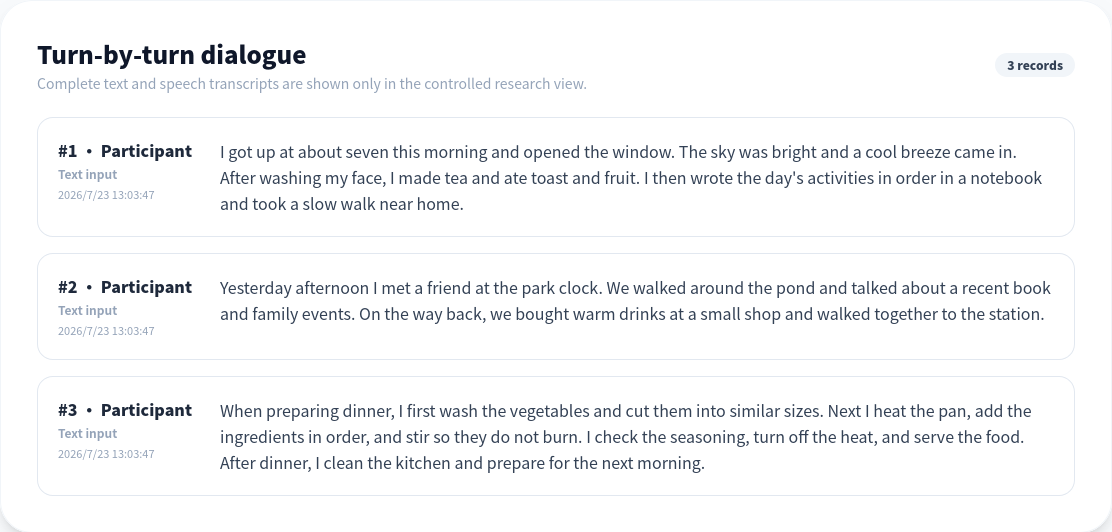}
  \caption{Ordered-dialogue view for the same synthetic demonstration session.
  The three participant turns appear in English translation, with their
  sequence, role, input mode, and timestamp.}
  \Description{A researcher-facing Turn-by-turn dialogue panel lists three
  synthetic participant records with sequence numbers, participant role, text
  input labels, timestamps, and the complete contributed text.}
  \label{fig:research-workbench-trace}
\end{figure*}

\FloatBarrier

\section{Technical Boundary-Check Details}
\label{app:technical-details}

\subsection{Fixed-Recording Manifest and Input Scope}

The fixed-recording audit used Cookie Theft picture-description sessions from
the original, non-noise-reduced English Pitt corpus in DementiaBank
\cite{becker1994natural}, pairing each recording with its CHAT transcript.\footnote{English Pitt corpus:
\url{https://talkbank.org/dementia/access/English/Pitt-orig.html}; corpus DOI:
\url{https://doi.org/10.21415/T5CS3J}. We acknowledge NIA grants AG03705
and AG05133, which support collection of the Pitt corpus.} Every run used the
same pre-existing input list as a fixed \emph{evaluation manifest}, not as a
train--test partition; this paper includes no task-specific training experiment.

Table~\ref{tab:dementiabank-manifest} summarizes the 109 sessions. The CHAT group field identifies 48 as
Control; the other 61 comprise 50 ProbableAD, four PossibleAD, and seven MCI
sessions. We collapsed the latter categories only for the binary evaluation and
report the targets neutrally as Corpus-C and Corpus-NC. The literal requested
output tokens are \texttt{Control} and \texttt{Impaired}; these reproduce the
source-corpus categories used for this evaluation.

\begin{table}[t]
  \caption{Composition of the fixed English evaluation manifest from CHAT
  metadata. Counts are recording-level except for participant IDs.}
  \label{tab:dementiabank-manifest}
  \centering
  \small
  \begin{tabular}{@{}lrrr@{}}
    \toprule
    Measure & All & Corpus-C & Corpus-NC \\
    \midrule
    Session recordings & 109 & 48 & 61 \\
    Participant IDs & 97 & 40 & 57 \\
    Age, median (range) & 69 (48--90) & 66 (48--80) & 74 (53--90) \\
    Gender field (F/M) & 75/34 & 32/16 & 43/18 \\
    Speech duration (min) & 123.35 & 48.41 & 74.94 \\
    \bottomrule
  \end{tabular}
\end{table}

The sessions carry 97 participant identifiers; 11 recur, but none crosses the
collapsed label. Metrics used the recording as the unit of analysis. The 109
readable MP3 files total 123.35 minutes and received one prediction from every
configuration. Text-only models received a cleaned transcript containing only
participant (\texttt{PAR}) turns, whereas speech models received the complete
session recording and speech-plus-text models received both. At least one
investigator (\texttt{INV}) turn is annotated in 106 transcripts. Median
investigator turns are three in Corpus-C and five in Corpus-NC; corresponding
median annotated investigator-speech durations are 2.1 and 5.9 seconds. The
speech branch can therefore expose interviewer behavior that the text branch
cannot, creating a possible shortcut rather than participant-only evidence.

\subsection{Fixed-Recording Model Comparison}
\label{app:fixed-comparison}

Table~\ref{tab:model-modality-comparison} in the main text reports accuracy
for all evaluated single-modality and multimodal configurations.

\emph{Zero-shot} denotes a fixed instruction without task-specific fine-tuning
examples, not guaranteed absence of DementiaBank from pretraining. Because
architecture and input scope both change, this is a failure-mode audit rather
than a controlled modality ablation or evidence that multimodality is superior.
Under the common open-model multimodal protocol, Qwen2-Audio and Ultravox
predicted Corpus-NC for all 109 recordings (100\% sensitivity and 0\%
specificity), and Step-Audio predicted Corpus-NC for 108 of 109 (100\%
sensitivity and 2.1\% specificity). Qwen2.5-Omni reached 62.4\% accuracy with
78.7\% sensitivity and 41.7\% specificity. Aggregate accuracy alone therefore
cannot determine interface content or production-model choice.

\subsection{Controlled Synthetic-Speech Construction}

We constructed 20 neutral older-adult personas from demographic attributes,
occupation, region, and ordinary life anchors. Persona specifications contain
no labels or cues about cognitive status or programmed difficulty. The set
comprises 10 women and 10 men, ages 66 to 87, and education from 10 to 18 years;
occupations and regions are unique. Ten personas received a morning-routine
prompt and ten an occupation-specific prompt.

For each persona, GPT-4o generated a fixed graph of six to nine underlying
events and rendered that same structure at four programmed
communication-difficulty levels. S0 is fluent, specific, and information dense;
S1 introduces occasional hesitation and vagueness; S2 adds frequent
word-finding difficulty, repetition, repair, filled pauses, and false starts;
and S3 adds pervasive hesitation, unresolved word finding, abandoned
utterances, loss of thread, and perseveration. Subject matter and intended
events were held within persona while linguistic realization changed.

We synthesized each rendering with \texttt{gpt-4o-mini-tts} using the
\texttt{coral} speech preset and level-specific delivery instructions. For
S1--S3, the
implementation split text at sentence boundaries and ellipses, synthesized
the segments, and concatenated them with locally inserted silences that
increased with level. The artifact contains 80 transcripts and 80 mono, 24-kHz
WAV files totaling 90.75 minutes.

\subsection{Synthetic Manipulation and Ordinal-Sensitivity Checks}

Surface linguistic markers were counted from verbatim transcripts and
normalized per 100 words. Silence ratio and pause count were recomputed from the
final WAV files with a frame-level RMS threshold; they are operational signal
measures rather than human-validated acoustic annotations.
Table~\ref{tab:synthetic-manipulation} shows how lexical diversity, disfluency,
and speech timing varied across the four programmed levels.

\begin{table*}[t]
  \caption{Mean (SD) linguistic and acoustic measurements across 20 personas
  at each programmed level. Fillers, ellipses, vague terms, and repair or
  lost-thread markers are rates per 100 words. TTR is type--token ratio; Dur. is
  duration in seconds.}
  \label{tab:synthetic-manipulation}
  \centering
  \scriptsize
  \setlength{\tabcolsep}{2.7pt}
  \begin{tabular}{@{}lrrrrrrrrr@{}}
    \toprule
    Level & Words & TTR & Fillers & Ellipses & Vague & Repair & Dur. &
    Silence & Pauses \\
    \midrule
    S0 & 116.70(11.94) & .681(.049) & 0.000(0.000) & 0.000(0.000) &
    .294(.486) & 0.000(0.000) & 40.23(4.81) & .313(.026) & 19.95(4.25) \\
    S1 & 104.40(11.20) & .652(.055) & 6.747(2.368) & 2.075(1.553) &
    .571(.707) & 0.000(0.000) & 46.37(5.58) & .394(.023) & 29.95(5.41) \\
    S2 & 108.55(11.10) & .576(.043) & 12.690(2.765) & 9.260(3.694) &
    1.553(1.551) & .277(.437) & 71.39(13.15) & .524(.044) & 48.50(9.98) \\
    S3 & 114.20(10.46) & .510(.048) & 20.083(4.446) & 21.393(6.179) &
    1.823(1.502) & .764(.892) & 114.26(22.95) & .646(.061) & 70.10(12.22) \\
    \bottomrule
  \end{tabular}
\end{table*}

Repeated-measures Friedman tests treated persona as the repeated unit. All eight
measures changed in the programmed direction ($p<.001$); Kendall's $W$ ranged
from .357 for vague terms to 1.000 for silence ratio. This supports internal
manipulation control, not correspondence to human impairment.

Qwen2.5-Omni-7B served as a sensitivity probe rather than a ground-truth
annotator. It completed 240 deterministic, zero-shot inferences: 20 personas by
four levels by three input modes. Each stimulus appeared in an independent
context and fixed randomized order; persona metadata and programmed level were
hidden. The prompt requested only a 0--3 score for observable communication
difficulty and prohibited conclusions beyond the stimulus.

Table~\ref{tab:synthetic-qwen-ordinal} in the main text reports the complete
ordinal-sensitivity results across the three input modes.

Every mode retained a non-decreasing four-level order for 19 of 20 personas,
but only text separated all four levels strictly for a substantial share of
personas (50\%; 0\% when speech was present). Speech scores averaged 1.05 points
below text (persona-bootstrap 95\% CI [-1.175, -.938]), and the probe never
assigned level 3 when speech was present. This may reflect the model's speech
pathway, the TTS realization, or their interaction. The same speech preset was
used throughout, S0 is unusually clean, and sex is confounded with task.

\subsection{Cross-Lingual Stress-Test Protocol}

We paired the 109 English cases with Chinese and Japanese translations that
preserve annotated pauses, repetitions, repairs, fragments, and nonverbal
events. CosyVoice3 rendered translated speech using an eight-second reference
window from each source recording \cite{du2025cosyvoice3}. This preserved case
identity and class balance while changing language, translation, and synthesis
together.

Qwen2.5-Omni-7B produced identical speech-plus-text accuracy for Chinese and
Japanese (62.39\%), while sensitivity/specificity differed at
52.46\%/75.00\% and 86.89\%/31.25\%, respectively. Speech-only accuracy was also
identical (57.80\%) despite Chinese sensitivity/specificity of
50.82\%/66.67\% and Japanese values of 96.72\%/8.33\%. The shift may arise from
translated disfluencies, synthesized prosody, model language priors, or their
interaction; the check cannot isolate a language effect.

\section{Participant-Facing Score Construction}
\label{app:score-construction}

The GPT-5.4 evidence-fusion component assigns each retained feature an integer
score $s_i$ from 0 to 100 in its predefined scoring direction through a fixed
structured prompt. The prompt defines 0 as no apparent matching pattern when
the source is available, 1--24 as minimal or isolated, 25--49 as mild, 50--74
as clear, and 75--100 as prominent or pervasive. Values between these anchors
are permitted, and each score must be accompanied by a source-specific
observation and confidence judgment. These anchors guide model output and are
not validated clinical thresholds. For the set $A$ of available features, the
trusted server computes a dimension or overall score as
\[
  S = \mathrm{round}\!\left(
    \frac{\sum_{i \in A} w_i s_i}{\sum_{i \in A} w_i}
  \right),
\]
where $w_i=1$ for a core feature and $w_i=0.5$ for a supportive feature. The
dimension calculation uses the available features within that dimension. The
overall index applies the same rule directly across all available retained
features rather than averaging the five dimension scores. Missing features are
excluded from both sums; if no feature is available, the index is not produced.

The weights encode the physician's ordinal priorities rather than measured
differences in predictive value. The fixed 2:1 ratio gives each core feature
twice the weight of each supportive feature. Alternative weighting ratios
were not evaluated.

The interface labels overall values from 0 to 33 as low, 34 to 66 as medium,
and 67 to 100 as high pattern prominence. These are descriptive display bands,
not validated clinical thresholds.

\section{Detailed Limitations}
\label{app:detailed-limitations}

The DementiaBank audit characterizes one fixed English Cookie Theft manifest,
not a random or population-matched sample. Its 109 sessions represent 97
participant identifiers, so repeated visits are not independent people; all
metrics are recording-level and cannot establish participant- or
population-level performance. Corpus-C and Corpus-NC also differ descriptively
in age, interviewer contribution, and some media formats. The text condition
contains participant turns only, whereas complete-session speech can contain
investigator speech; differences between modality configurations therefore
cannot be interpreted causally. Zero-shot describes our prompt-only procedure,
not model pretraining, and we cannot rule out prior DementiaBank exposure.

\MindBridge{}'s clinical grounding comes from multi-stage involvement by one
psychiatry resident physician: an initial consultation, 33 structured
judgments over the 11-family inventory, review of labeled sample speech,
written rationales governing feature priority and disclosure, and review of
the prototype UI. It does not constitute multi-clinician consensus,
participatory co-design, or clinical validation. We assembled
the inventory, organized retained families into five dimensions, and converted
categorical judgments into numeric weights. The resulting 0--100 research
signal is neither standardized nor norm referenced and cannot establish
overall cognitive ability.

The user study used a small mixed-age convenience
sample of ten, with only three participants in age bands of 60 or older and
uneven first exposure to the score-only report (Package A, $n=2$), the
five-dimension report (Package B, $n=5$), and the interpretation-first report
(Package C, $n=3$). P9 viewed Package B for the high-index case (Case H) and
Package C for the low-index case (Case L), whereas the other nine participants
retained one package across both cases. The sample cannot
estimate an age effect, population prevalence, or package superiority. The
disclosed one-sample boundary, fixed order of Case H followed by Case L, heterogeneous P9 exposure, and
later comparison of all packages introduce instruction and transfer. First
attention and accommodation were self-reported; the study-specific items are not
validated scales; open-response codes were not independently double-coded. The
two complementary tasks used one passage and open narrative prompt, and their
ratings evaluated interaction rather than memory accuracy. The export lacks
recruitment, facilitation, device, active-duration, and build-version metadata.
Typing does not evaluate microphone control, ASR, spoken timing, acoustic
features, or speech-interaction usability.

The synthetic extension used generated language and speech rather than
independently characterized human speech. The 20-persona batch used one TTS
speech preset, contains a task--sex confound, and implemented a fixed one-turn
protocol.
Generator--evaluator circularity remains possible even with different models.
The cross-lingual check combined translation and speech synthesis, so it cannot
isolate a language effect or estimate performance on naturally produced
Chinese or Japanese speech. These constraints define the evidence needed in
subsequent evaluations of naturally produced multilingual speech and real-world
use.

\end{document}